\documentclass[]{aa}

\usepackage{times,epsfig,amssymb,amsmath,natbib}
\newcommand{\chandra}{{Figures/\it Chandra }}

\usepackage{tabularx} 
\usepackage{footmisc}

\begin{document}
\title{On the evolution of the entropy and pressure profiles \\
in X-ray luminous galaxy clusters at $z > 0.4$}
\titlerunning{Entropy and pressure profiles in X-ray galaxy clusters at $z > 0.4$}

\author{V. Ghirardini\inst{1,2} \and  S. Ettori\inst{2,3} \and S. Amodeo\inst{4} \and R. Capasso\inst{	5,6} \and M. Sereno\inst{1,2}}
\authorrunning{V. Ghirardini et al.}

\institute{
 Dipartimento di Fisica e Astronomia Universit\`a di Bologna, Via Piero Gobetti, 93/2, 40129 Bologna, Italy
 \and INAF, Osservatorio Astronomico di Bologna, Via Piero Gobetti, 93/3, 40129 Bologna, Italy
 \and INFN, Sezione di Bologna, viale Berti Pichat 6/2, I-40127 Bologna, Italy
 \and LERMA, Observatoire de Paris, PSL Research University, CNRS, Sorbonne Universit\'es, UPMC Univ. Paris 06, F-75014 Paris, France 
 \and Department of Physics, Ludwig-Maximilians-Universitaet, Scheinerstr. 1, 81679 Muenchen, Germany
 \and Excellence Cluster Universe, Boltzmannstr. 2, 85748 Garching, Germany
}

\offprints{V. Ghirardini}
\mail{vittorio.ghirardini2@unibo.it}
\abstract
{Galaxy clusters are the most recent products of hierarchical accretion over cosmological scales. The gas accreted from the cosmic field is thermalized inside the cluster halo. Gas entropy and pressure are expected to have a self-similar behaviour with their radial distribution following a power law and a generalized Navarro-Frenk-White profile, respectively. This has been shown also in many different hydrodynamical simulations. 
}
{We derive the spatially-resolved thermodynamical properties of 47 X-ray galaxy clusters observed with Chandra in the redshift range 0.4 $<$ z $<$ 1.2,
one of the largest sample investigated so far with X-ray spectroscopy and with mass reconstructed using hydrostatic equilibrium equation, 
with a particular care in reconstructing the gas entropy and pressure radial profiles.
We search for deviation from the self-similar behaviour and look for possible evolution with redshift.
}
{
Under the assumption of a spherically symmetric distribution of the intracluster plasma, we combine the deprojected gas density and the deprojected spectral temperature profiles via the hydrostatic equilibrium equation in order to constrain the concentration and the scale radius, which are the parameters that describe a Navarro-Frenk-White profile for each of the cluster in our sample. The temperature profile, that combined with the observed gas density profile reproduces the best-fit mass model, is then used to reconstruct the profiles of the entropy and pressure. 
These profiles cover a median radial interval of  [0.04 $R_{500}$ -- 0.76 $R_{500}$]. After interpolating on the same radial grid and partially extrapolating up to $R_{500}$, these profiles 
are then stacked in order to increase the precision of the analysis, also in 3 independent redshift bins.
The gas mass fraction is then used in order to improve the self-similar behaviour of the profiles, by reducing the scatter among the profiles by a factor 3.}
{The entropy and pressure profiles lie very close to the baseline prediction from gravitational structure formation. 
We show that these profiles deviate from the baseline prediction as function of redshift, in particular at $z>0.75$,
where, in the central regions, we observe higher values of the entropy (by a factor of $\sim 2.2$) and systematically lower estimates (by a factor of $\sim 2.5$) of the pressure.
The effective polytropic index, which retains informations about the thermal distribution of the gas, shows a slight linear positive evolution with the redshift and the concentration of the dark matter distribution.
A prevalence of non-cool-core, disturbed systems, as we observe at higher redshifts, can explain such behaviours.
}
{}

\keywords{Galaxies: clusters: intracluster medium -- Galaxies: clusters: general -- X-rays: galaxies: clusters -- (Galaxies:) intergalactic medium }

\maketitle
\begin{figure*}[t]
\hbox{
 \includegraphics[width=0.5\textwidth]{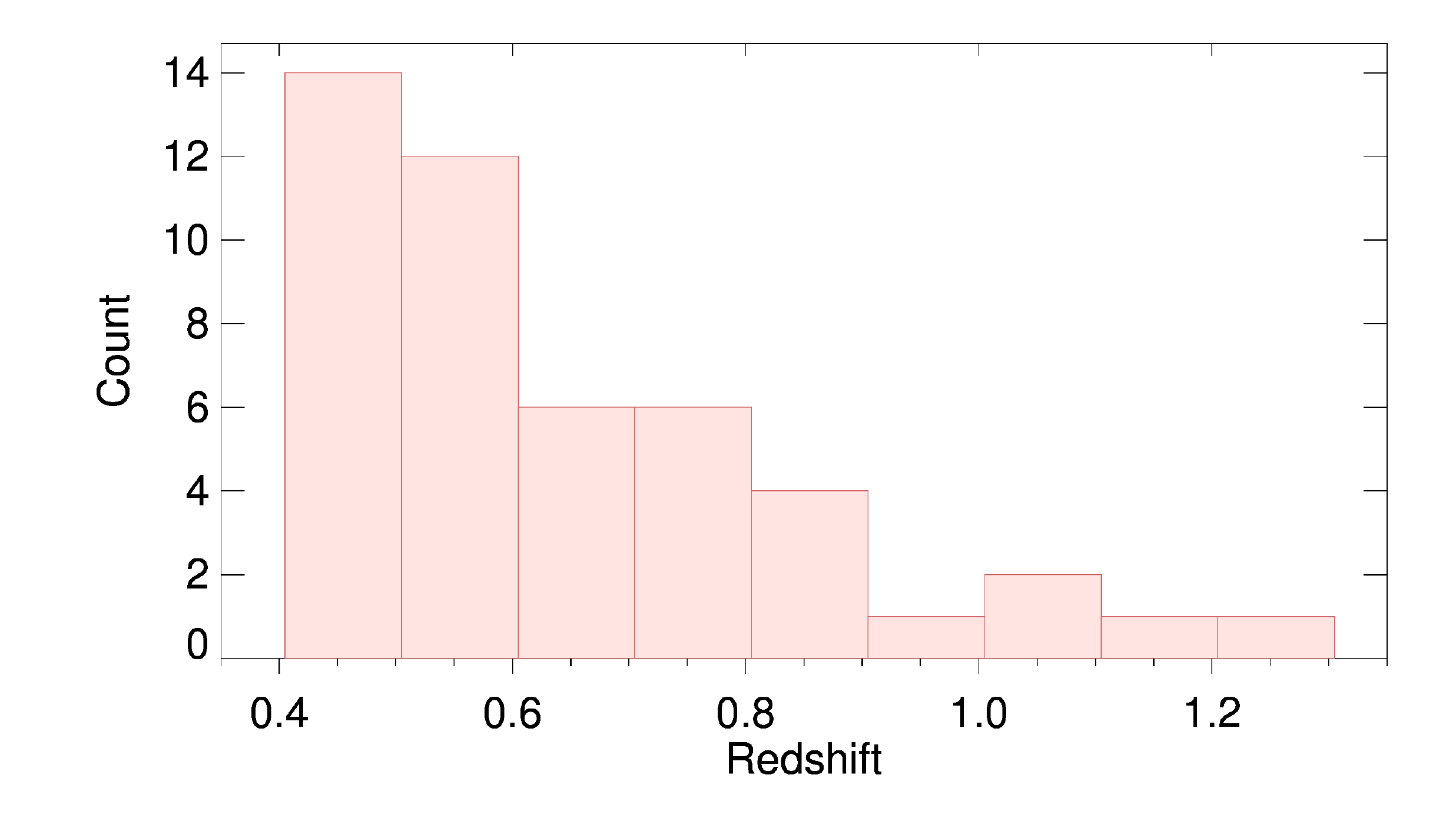}
 \includegraphics[width=0.5\textwidth]{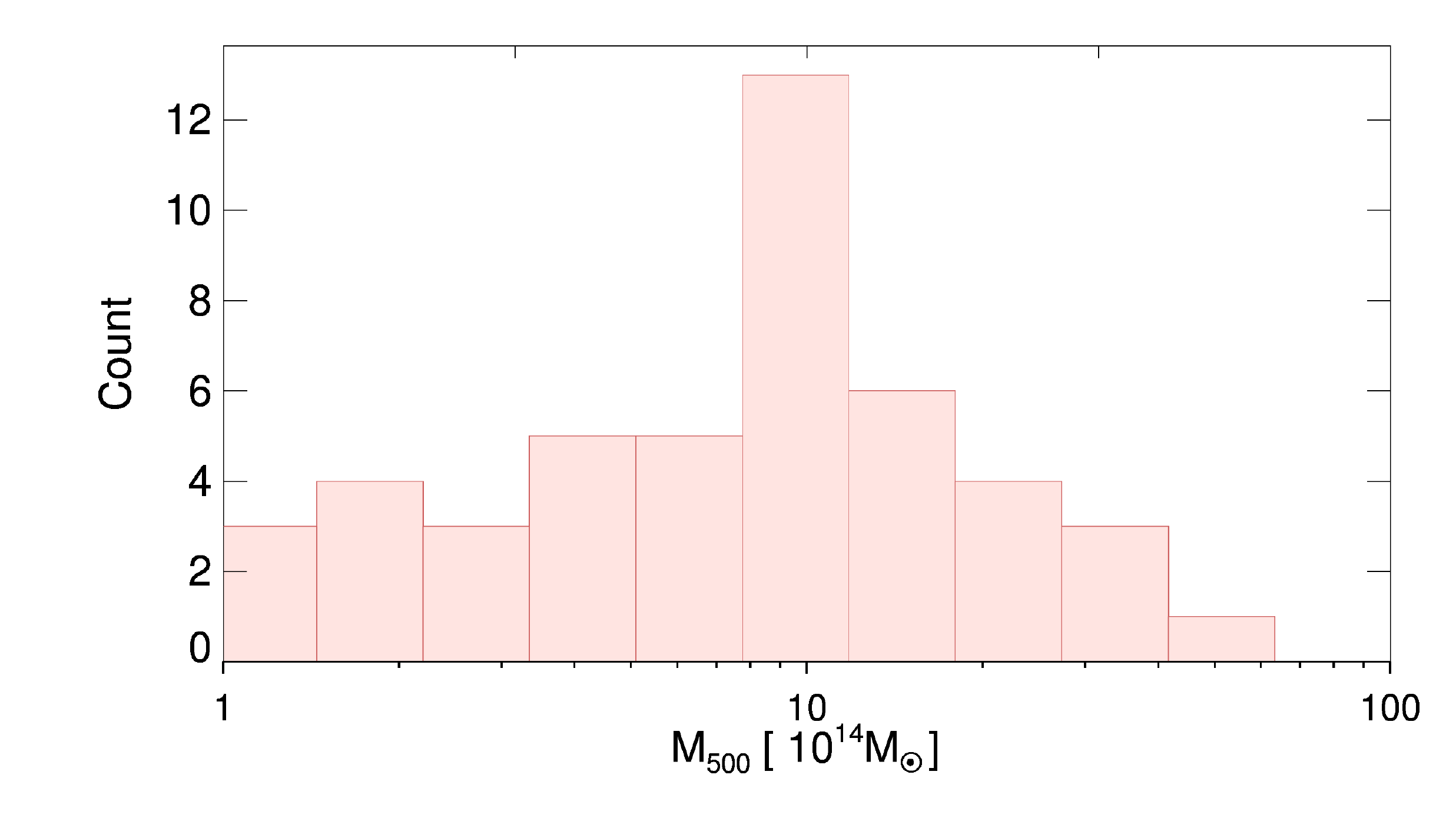}
 }
\caption{Redshift (left) and total mass (right) distribution of the clusters in our sample. }
\label{sample}
\end{figure*}

\section{Introduction}
Cosmic structures evolve hierarchically from the primordial density fluctuations, growing to form larger and larger systems under the action of gravity. 
Clusters of galaxies are the biggest virialized structures in the Universe and aggregate as bound objects at a relatively late time ($z<3$). 
The cosmic baryons fall into the gravitational potential of the cold dark matter (CDM) halo and under the action of the collapse 

and subsequent shocks, adiabatic compression and turbulence, heat up to the virial temperature of few keV and form a fully ionized X-ray emitting intra-cluster medium (ICM) \citep{tozzi+01,voit+05,zhuravleva+14},

Modelling the ICM emission by thermal brehmsstrahlung allows observations in the X-ray band to provide a direct probe of the gas (electron) density, $n_{\rm e}$, and gas temperature, $T$. 
Assuming the perfect gas law and a monoatomic gas, 
the pressure is recovered as $P = T \ n_{\rm e}$ and the specific entropy as $K = T / n_{\rm e}^{2/3}$.

Entropy is a fundamental quantity to track the thermal history of a cluster, since it always rises when heat is produced \citep[see][]{voit+05}. In the presence of non-radiative processes only, low-entropy gas would sink to the centre of the cluster while high-entropy gas would expand, producing the typical self-similar radial distribution
which follows a power law with a characteristic slope of 1.1 \citep[e.g.][]{tozzi+01}. Deviations from this predicted behaviour have been observed in the central region of clusters \citep[see][]{ponman+99}.  Simulations show that non-gravitational cooling and heating processes, such as radiative cooling and subsequent  AGN and supernovae feedback, break self-similarity in the inner region of galaxy clusters \citep[e.g.][]{voit+05,mccarthy+17}. 
To justify the observed deviations from self-similarity, \cite{tozzi+01} propose a model where an entropy {\it floor}
is present before the gas is accreted by the dark matter halo.

Only in the last few years some dedicated papers have studied the evolution of the entropy profiles with redshift. \cite{mcdonald+13},  \cite{mcdonald+14}, and more recently  \cite{bartalucci+16}, conclude that both cool core clusters (CC) and non cool core clusters (NCC) in samples selected through their Sunyaev-Zeldovich \citep[][hereafter SZ]{sz80} signal have similar entropy profiles, with no relevant changes with the cosmic time.
 
On the other hand, pressure is the quantity the least affected by the dynamical history and non-gravitational physics \citep[]{arnaud+10}. Using the model introduced by \cite{nagai+07}, \cite{arnaud+10} show that pressure has a very regular radial profile with only small deviations about the mean (less than 30 per cent outside the core regions). This is confirmed by recent observations of SZ-selected clusters \citep[e.g.][]{mcdonald+14,bartalucci+16} showing no significant deviation from the ``universal profile'' within $R_{500}$ up to redshift z=0.6. 

In this work, we use the analysis on the mass distribution of 47 galaxy clusters in the redshift range 0.405--1.235 presented in \cite{amodeo+16} in order to study the radial shape of pressure and entropy at different redshifts, looking for deviations from the self-similar behaviour and for evolution with cosmic time.

In the present study, we assume a flat $\Lambda CDM$ cosmology, with matter density parameter $\Omega_M = 0.3$ and an Hubble constant of $H_0=70 $ km s$^{-1}$ Mpc$^{-1}$.

Therefore the critical density of the universe is:
\begin{equation}
\rho_{c} \equiv \frac{3 H^2(z)}{8 \pi G} = \frac{3 H_0^2}{8 \pi G} E^2(z) \approx 136 \frac{M_\odot}{kpc^3} E^2(z)
\end{equation}
where $E(z) \equiv H(z)/H_0 = \big[ \Omega_M (1+z)^3 + \Omega_\Lambda \big]^{1/2}$.

In our study, we consider also a rescaling dependent on the halo's mass. To apply this, we measure the quantities of interest over the cluster's regions defined by an overdensity $\Delta$, defined as a region for which the mean mass density is $\Delta$ times the critical density of the universe.

In the following analysis, we choose $\Delta= 500$, considering that our profiles have a radial extent of the same order of magnitude as $R_{500}$ (see Fig. \ref{fig:maxT}).
By definition, $M_{500}$ is then equal to $4/3 \pi \, 500 \rho_{c} \, R_{500}^3$.

All the quoted statistical uncertainties are at $1 \sigma$ level of confidence.

All the fitting processes are, unless otherwise stated, executed using the IDL routine MPFIT \citep{mpfit}, which performs a Levenberg-Marquardt least-squares fit weighting with both the errors on x and y axes.

The paper is organized as follows: in Section~2 we present the sample, in Section~3 we present the method we apply to reconstruct the entropy and pressure profiles.
The data analysis is detailed in Section~4 , with an exhaustive discussion of our results presented in Section~5. 
We draw our conclusions in Section~6.

\begin{figure}[t]
\includegraphics[width=0.5\textwidth]{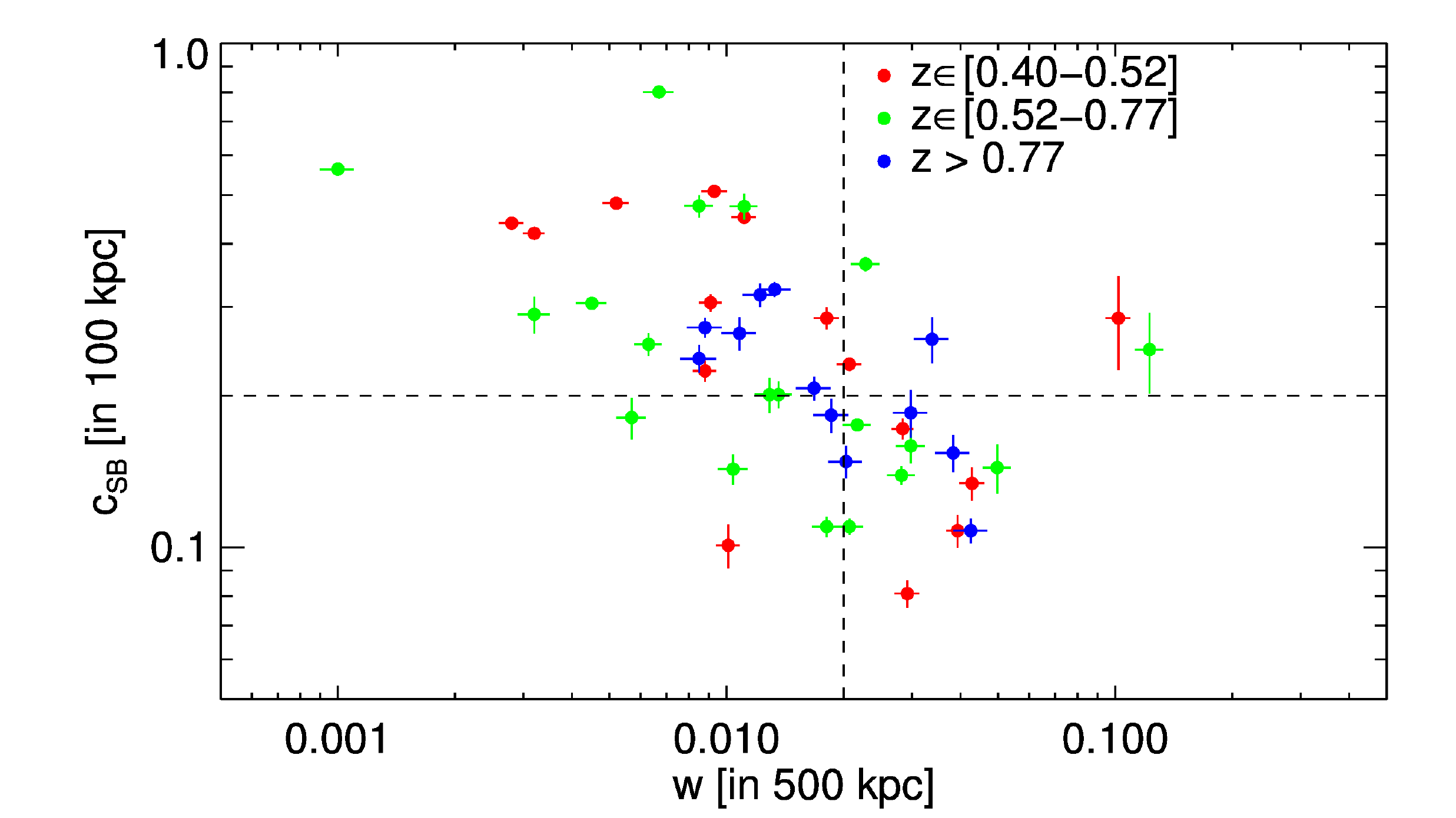}
\caption{Morphological parameters distribution in the plane of $w$ versus concentration $c_{SB}$. }
\label{Figure:morph}
\end{figure}
\section{Sample properties}
The sample selection and X-ray analysis is thoroughly explained in \cite{amodeo+16}. In this work, however, we will recall some of the most important properties.
Being massive objects, the relatively long \chandra exposure time ($t_{exp} >$  20 ksec) considered for each cluster permits 
to extract at least 3 independent spectra over the cluster's emission which allows to make a complete X-ray analysis as it is usually done for  low redshift clusters.  \citep[see][for details on the sample selection and X-ray analysis]{amodeo+16}.
We point point out that clusters which are undergoing a major merger are excluded from the analysis since they would strongly break the hydrostatic equilibrium assumption, which is essential in recovering the mass profile.
A completeness analysis of the sample has been made \citep[see][]{amodeo+16}: the selection criteria have effectively choosen the very massive high end of the cluster halo function in the investigated redshift range.

In Fig.~\ref{sample}, we show the redshift and mass distributions of the objects in our sample, with 9 systems at $z>0.8$ and 8 with an estimated $M_{500}$ larger than $2 \times 10^{15} M_{\odot}$.

We use the centroid shift $w$ and the concentration $c_{SB}$ \citep[both resulting from the analysis made in ][]{amodeo+16} to distinguish clusters with a cool core and clusters without it: we follow the results of \cite{cassano+10}, choosing clusters with $w < 0.012$ and $c_{SB} > 0.2$ to be the classical relaxed CC. These objects are located in the upper-left quadrant of Fig.~\ref{Figure:morph}. The main goal of this paper is to study the evolution of entropy and pressure with cosmic time, therefore the sample is divided in redshift bins. Nevertheless a parallel analysis of the cluster sample has been made, where clusters are divided in CC and NCC based on their morphological properties, using the criteria found by \cite{cassano+10} mentioned above.

In our sample, the gas density profiles, obtained from the geometrical deprojection of the observed surface brightness profiles, cover the median radial range of [0.04 $R_{500}$ -- 0.76 $R_{500}$], with a mean relative error of 21\%.
In Fig.~\ref{fig:maxT}, we show the observed distribution of the minimum and maximum radius in the gas density profiles. 
We point out that half of the clusters have a radial coverage that extents above 0.77 R$_{500}$.

\section{The method to reconstruct $K(r)$ and $P(r)$}

\cite{amodeo+16} presented the method applied to constrain the mass distribution of the galaxy clusters in our sample under the assumption that a spherically symmetric ICM is in hydrostatic equilibrium with the underlying dark matter potential. The backward method adopted \citep[see][]{ettori+13} allows to constrain the parameters of a mass model, i.e. the concentration and the scale radius for a NFW model \cite[][]{nfw+97}, using both the gas density profile, obtained from the geometrical deprojection of the X-ray surface brightness profile, and the spatially resolved spectroscopic measurements of the gas temperature. 
As by-product of the best-fit mass model, we obtain the 3D temperature profile associated to the gas density measured in each radial bin.
In other words, we obtain an estimate of the ICM temperature in each volume’s shell where a gas density is measured from the geometrical deprojection of the X-ray surface brightness profile, in such a way that, inserting the temperature and density profiles into the hydrostatic equilibrium equation \citep{hee}, the best-fit mass model is reproduced. From the combination of these profiles, thermodynamical properties like pressure and entropy are recovered.

Following \cite{voit05}, temperature, pressure and entropy associated to this halo's overdensity are, respectively:
\begin{flushleft}
\begin{equation}
k_B T_{500} =  10.3 \text{\ keV} \bigg(\frac{M_{500}}{1.43 \cdot 10^{15} M_\odot} \bigg)^\frac{2}{3} E(z)^{2/3} \hfill
\label{eq:T500}
\end{equation}
\begin{equation}
P_{500} =  1.65 \cdot 10^{-3} \text{\ keV} \text{\ cm}^{-3} \bigg(\frac{M_{500}}{3 \cdot 10^{14} M_\odot}\bigg)^{2/3} E(z)^{8/3}
\label{eq:P500}
\end{equation}
\begin{equation}
K_{500} =  103.4 \text{\ keV cm}^2 \bigg(\frac{M_{500}}{10^{14} M_\odot}\bigg)^{2/3} E(z)^{-2/3} f_{b}^{-2/3} 
\label{eq:K500}
\end{equation}
\begin{equation}
K(R)/K_{500} =  1.42 \; \left(\frac{R}{R_{500}} \right)^{1.1}
\label{eq:voit}
\end{equation}
\end{flushleft}
where Eq. \ref{eq:voit} is the \cite{voit+05} prediction where the radial dependence of $K$ has been here rescaled from $\Delta=200$ to $500$ using 
the ratio $\frac{R_{500}}{R_{200}} = 0.66$ as predicted from a NFW mass model with $c_{200}=4$, typical for massive systems.

\begin{figure}[t]
\includegraphics[width=0.5\textwidth]{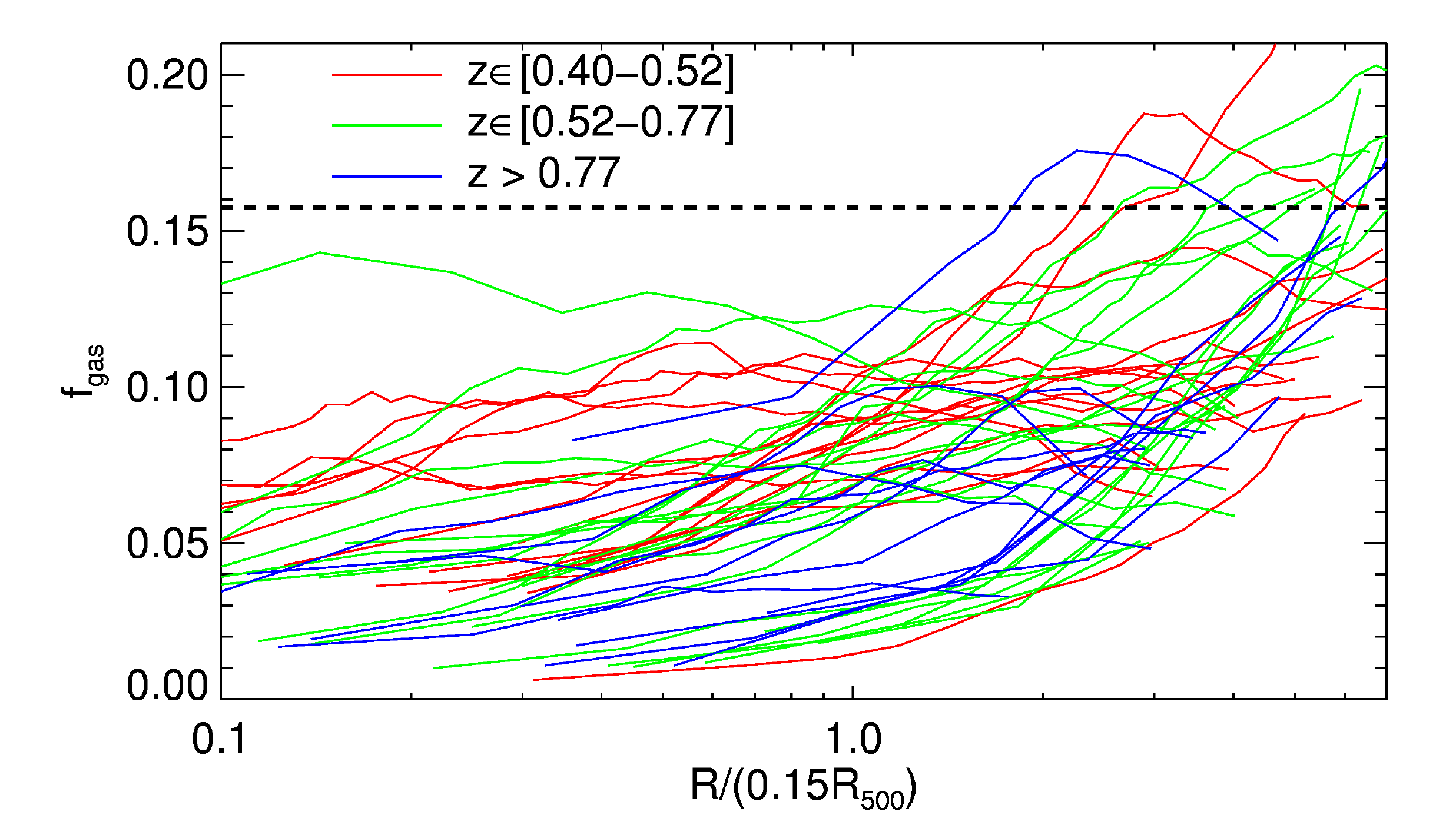}
\caption{Gas fraction profile for all our clusters. The black dashed line represents the universal baryon fraction from {\it Planck} \citep{planck+16}}
\label{Figure:f_gas}
\end{figure}

The importance of using a proper rescaling which includes the gas mass fraction to reach self-similarity in the entropy profiles was pointed out in the work of \cite{pratt+10}.
The gas fraction, $f_{gas}(<r) = M_{gas}(<r) / M_{tot}(<r)$, is here defined as the ratio between the gas mass obtained from the integration of the gas density over the cluster's volume and the hydrostatic mass, and is thus reconstructed directly from our data. We show the radial profiles of the gas fraction in Fig.~\ref{Figure:f_gas}. Throughout this paper we will consider the universal baryon fraction to be $f_b=0.15$ \citep{planck+16}. 
In Table~\ref{table:extra} we list additional information not presented in \cite{amodeo+16}, like references on the redshifts and the gas mass fraction at $R_{500}$.

\section{Data Analysis}
\label{ana}

Because of the poor statistics of each profile, we combine them.
In order to have at least one point in each radial bin, chosen logarithmically with total number of bins equal to 30\footnote{The number of bins equal to 30 has been chosen to guarantee a proper radial coverage}, over the range $[0.015 R_{500}, R_{500}]$, first, we have extrapolated the data using the best-fit power law plus a constant for the entropy profile,
and the functional form introduced by \cite{nagai+07} for the pressure profile:
\begin{equation}
\frac{P(x)}{P_{500}} = \frac{P_0}{(c_{500}x)^\gamma [1+(c_{500}x)^\alpha]^{\frac{\beta-\gamma}{\alpha}}}
\label{eq:arnaud}
\end{equation}
where $x=R/R_{500}$ and $\gamma$, $\alpha$ and $\beta$ are respectively the central slope, the intermediate slope and the outer slope defined by a scale parameter $r_s = R_{500}/c_{500}$ ($R << r_s$, $R \sim r_s$ and  $R >> r_s$ respectively). All the parameters of this function are left free except for the outer slope $\beta$ and the inner slope $\gamma$, which is fixed to the ``universal'' values\citep[5.49 and 0.308, ][]{arnaud+10}.

We have calculated the value of the thermodynamical quantities in the extrapolated radial points, using the functional forms indicated above. The error associated to the thermodynamic quantities in the extrapolated radial points is the sum in quadrature of the propagated best-fitting parameters and the median uncertainty estimated in the last five radial bins of the raw profiles.

Furthermore, to investigate the average behaviour of these profiles as function of redshift,
we divided the dataset into 3 redshift bins, chosen in order to have approximatively the same number of clusters so that the resulting profiles will have an approximatively constant signal to noise ratio: 
15 with $z \in [0.4, 0.52]$; 20 with $z \in [0.52, 0.77]$; 12 with $z > 0.77$. 
In each redshift bin, the profiles are stacked in logarithmic space using the inverse of the 1$\sigma$ error as weights, meaning that at each radial point the weighted mean is:
\begin{equation*}
<x> = \frac{\Sigma w_i x_i}{\Sigma w_i} \text{ with } w_i = \sigma_i^{-2}
\end{equation*}
where $x$ is the logarithm of the considered thermodynamic quantities (pressure or entropy).

This ``stacking'' procedure ensures a higher statistical significance of the measured ``mean'' quantities, which propagates into a lower uncertainties in constraining the best-fit parameters of the functional forms adopted.

\begin{figure}[ht!]
\includegraphics[width=0.48\textwidth]{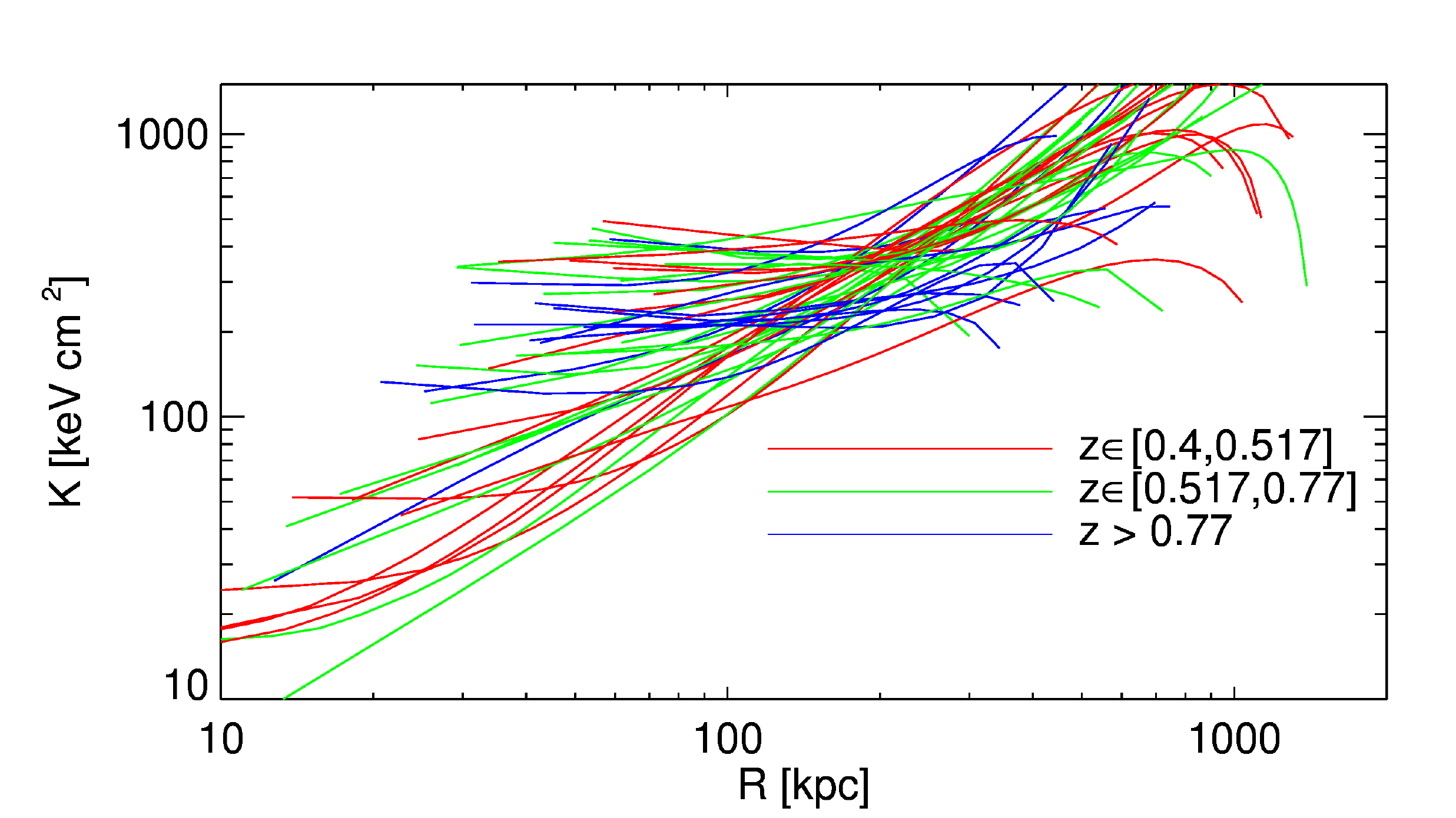}
\includegraphics[width=0.48\textwidth]{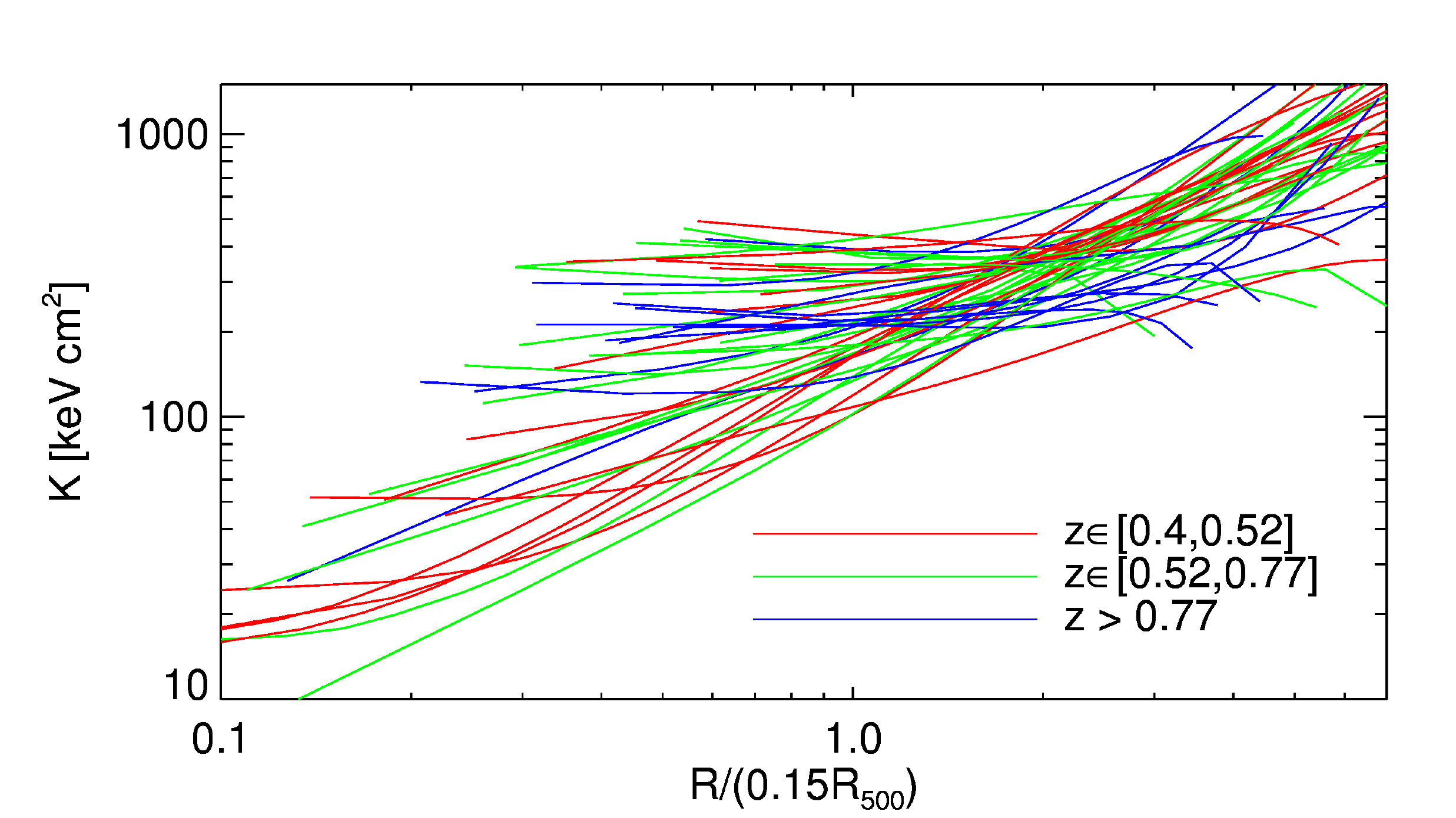}
\includegraphics[width=0.48\textwidth]{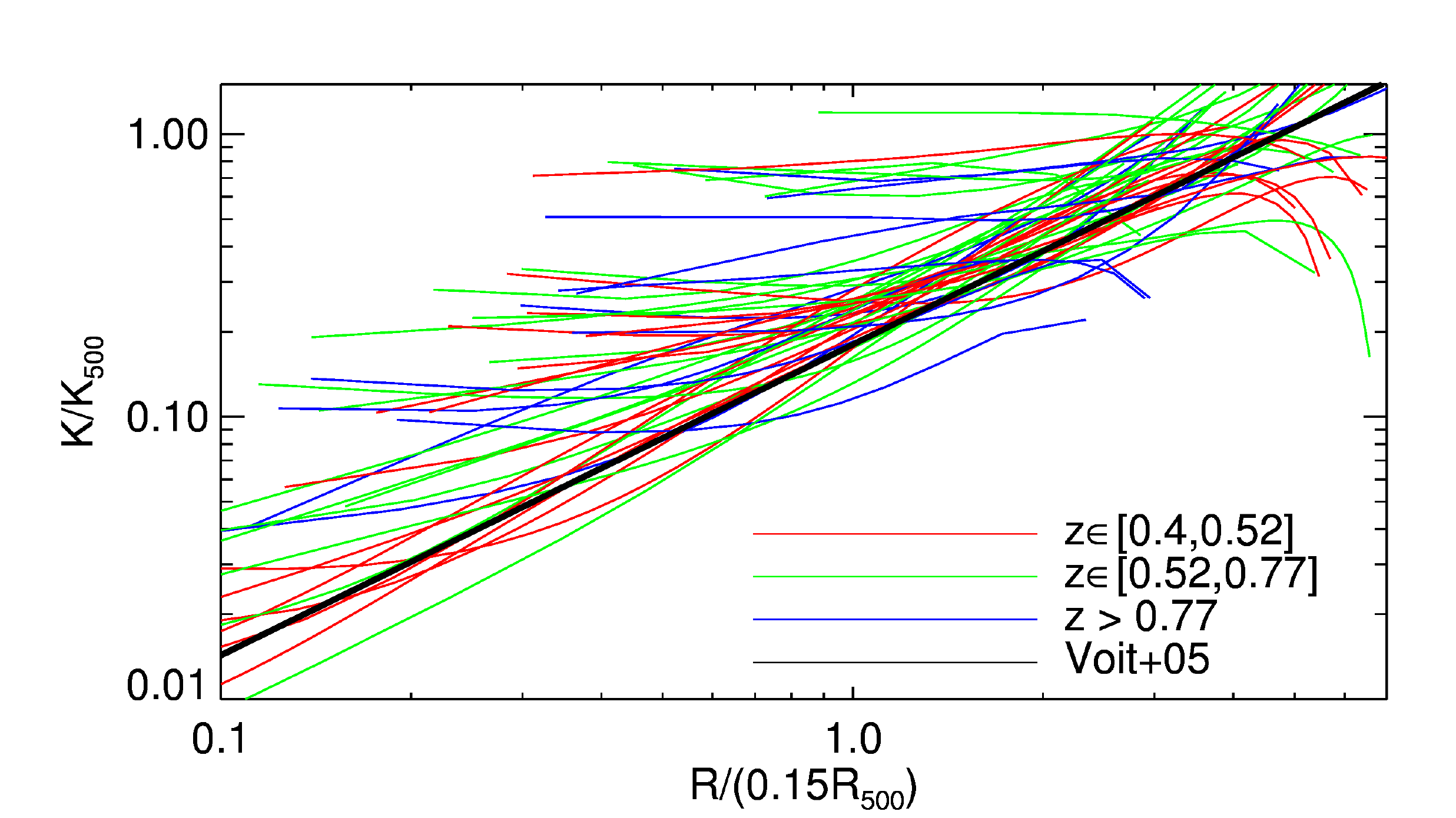}
\includegraphics[width=0.48\textwidth]{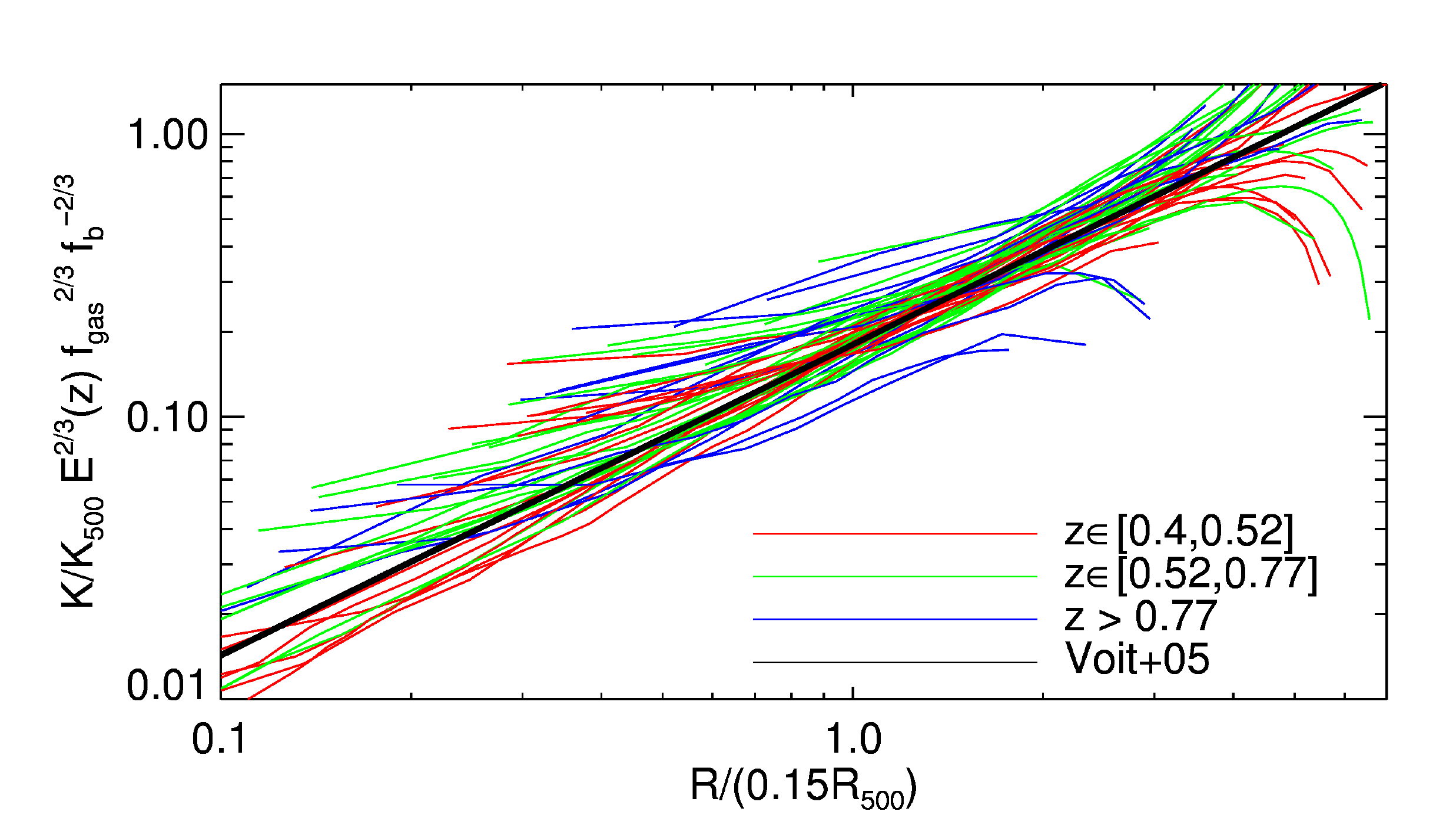}
\caption{The complete sample of our data has been plotted color coded with respect to redshift. We can see the effect of rescaling from top to bottom, from largely scattered data to coherent ones.
No error bars are drawn for sake of clarity. }
\label{fig:rescaling}
\end{figure}

\subsection{Fitting procedure for the entropy profile}

In Fig.~\ref{fig:alldata}, we show the reconstructed entropy profiles, rescaled as described below using $K_{500}$, of the 47 clusters in our sample prior to the application of the extrapolation and stacking procedure.

The radial behaviour of the entropy distribution is commonly described with a power law plus a constant term accounting for the combined action of cooling and heating feedbacks which affect the central regions \citep[see][]{cavagnolo+09}.
\begin{equation}
K=K_{0}+K_{100} \Big(\frac{R}{100 kpc} \Big)^\alpha
\label{eq:k0_k100_alpha}
\end{equation}
This functional form has an underlaying physical sense when we rescale with 100 kpc, since typically deviations from non-radiative simulations are seen below this radius, where cool core clusters and non cool core clusters actually differ. $K_0$ is the central entropy, which has been used in several work \citep[i.e.][]{cavagnolo+09,voit+05,mcdonald+13} to discriminate between relaxed CC clusters, with $K_0 \sim$ 30 keV cm$^2$, and disturbed NCC, with $K_0 >$ 70 keV cm$^2$.

We  also consider a functional form where the scaling is done with respect to $0.15 R_{500}$ in order to take into account the dimension of the core in systems at different mass and redshift:
\begin{equation}
\frac{K}{K_{500}}=K'_{0}+K'_{0.15} \Big(\frac{R}{0.15 R_{500}} \Big)^{\alpha'}.
\label{eq:abc}
\end{equation}

As shown in \cite{pratt+10}, we expect the scatter among clusters' entropy profiles to be suppressed even more when the renormalization includes 
both the global and the radial dependence on the gas mass fraction:
\begin{equation}
E(z)^{2/3} \frac{K}{K_{500}} \Bigg(\frac{f_{gas}}{f_{b}} \Bigg)^{2/3}=K''_{0}+K''_{0.15} \Big(\frac{R}{0.15 R_{500}} \Big)^{\alpha''},
\label{eq:abcd}
\end{equation}

In Fig.~\ref{fig:rescaling} we show how our refinement in the rescaling procedure of the entropy profiles, from ``no rescaling'' at all to the inclusion of the dependence upon the gas fraction (see Equation~\ref{eq:abcd}), reduces the scatter among the profiles, improving the agreement with the self-similar prediction \citep{voit+05}.

\subsection{No rescaling}

In the first panel of Fig.~\ref{fig:rescaling} we plot all the entropy profiles without applying any rescaling, i.e. the actual physical size in kpc versus entropy in keV cm$^2$. We name them as ``raw data'', because the physical values are reported without any rescaling. We can deduce that our clusters have very different thermodynamic histories, in fact at each radial point the profiles spans more than one order of magnitude, which exclude self-similarity.

We fit all the entropy profiles using Equation \ref{eq:k0_k100_alpha}, in order to look at the occupation of the parameter space in our sample.
For each cluster we obtain a value for each one of the parameters $K_0$, $K_{100}$ and $\alpha$ and we plot all those values in Fig.~\ref{fig:k0_k100_alpha}. For each one of the three histograms we fitted with a lognormal distribution. We dedicated a special attention to the parameter $K_0$, since it was shown, in the work of \cite{cavagnolo+09}, that central entropy may have a bimodal distribution. 
We have got $\sim$10 clusters with $K_0 = 0$ and $\sim$3 with $\alpha = 0$.  
 
In the central regions of galaxy clusters, no significant emission from gas at very low temperature is observed \citep[e.g.][]{peterson+06}, limiting the central value of the gas entropy to be positive ($K_0>0$). 
For this reason these points are excluded from Fig.~\ref{fig:k0_k100_alpha}.

The parameter $K_0$ (top panel) has not a clear bimodal distribution. The data exibit a peak between 150 and 200 keV cm$^2$ and a significative tail for low values of $K_0$.   The best fit obtained using two lognormal distributions shows one peak at 130 keV cm$^2$ and one at 5 keV cm$^2$, with $\chi^2_{red} = 1.83$, while the unimodal fits yields a peak at 100 keV cm$^2$ with $\chi^2_{red} = 2.03$. 
We used the Bayesian information criterion \citep[BIC,][]{bic} to distinguish if there are statistical differences between the two fitted models: a $\Delta BIC$ between 2 and 6 indicates positive evidence against the model with higher BIC, while at values greater than 6 the evidence is strong. We obtained $\Delta BIC = 0.02$, and therefore we can't discriminate between a unimodal and a bimodal distribution.
Even tough the best fit centers are close to the values found by \cite{cavagnolo+09}.

\begin{figure}[t]
\includegraphics[width=0.5\textwidth]{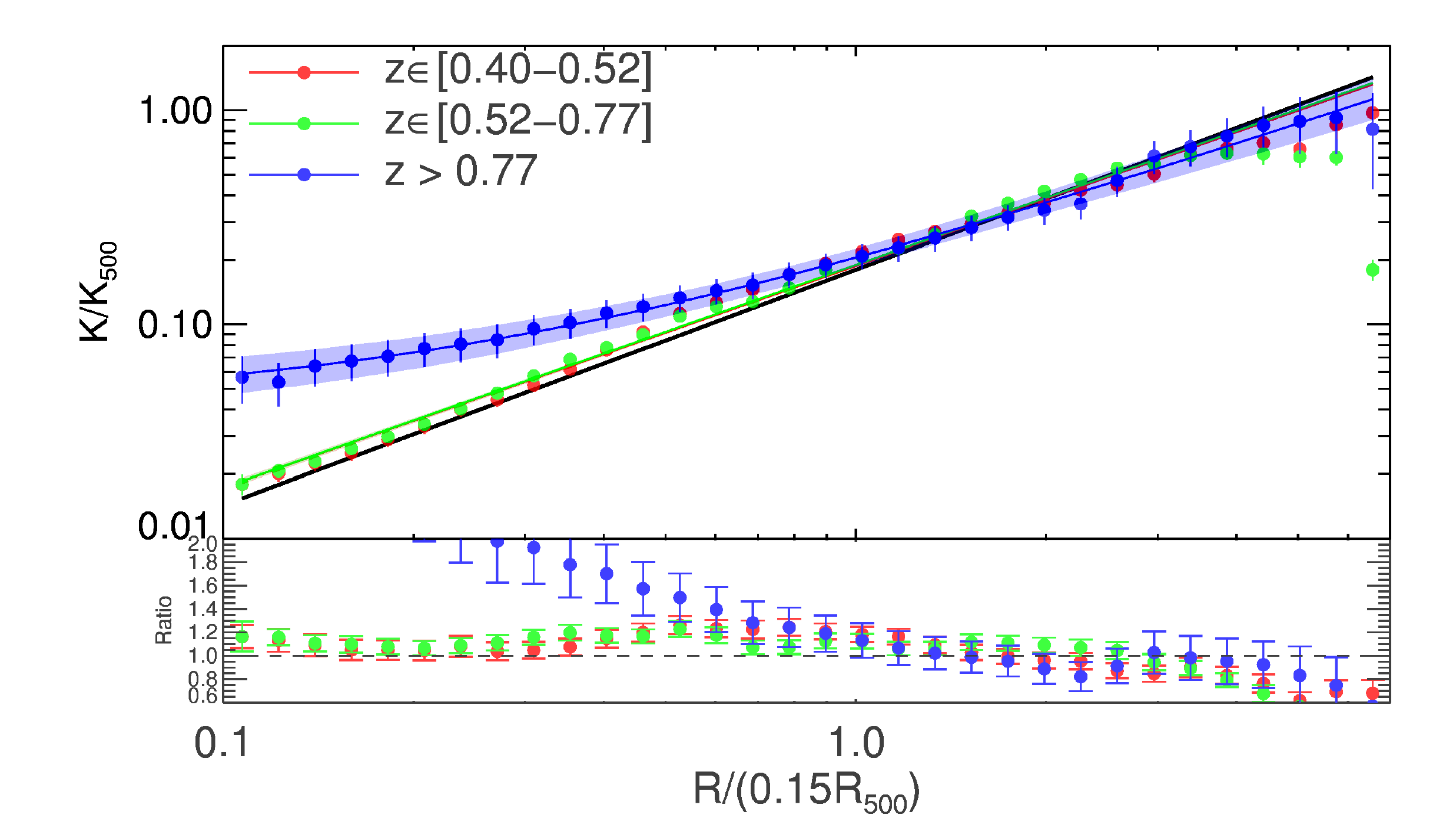}
\caption{Stacked entropy profiles and fits in the 3 redshift bins. The shaded areas represents the errors on the fit. 
The colors of the points are representative of the redshift bin considered.
 The black line represents the non-radiative prediction \citep{voit+05}}
\label{Figure:K500_errorevolution}
\end{figure}

The distribution of the parameter $K_{100}$ (middle panel of Fig.~\ref{fig:k0_k100_alpha}) is well fitted by a lognormal distribution with median value corresponding to $104\pm^{170}_{64}$\footnotemark[2] keV cm$^2$  and $\chi^2_{red} = 0.17$. The distribution of this parameter is smooth without any significative tail, this indicates that fitting considering 100 kpc as the typical core dimension generates a well defined distribution for $K_{100}$ resembling a lognormal one.

The distribution of the power law index $\alpha$ (bottom panel of Fig.~\ref{fig:k0_k100_alpha}) is, as $K_{100}$, well fitted by a lognormal distribution with median value $1.40\pm^{0.67}_{0.45}$\footnotemark[2] and $\chi^2_{red} = 1.1$. 
The visible peak has a higher value than the non-radiative prediction \citep{voit+05}, but nevertheless it is compatible within 1$\sigma$.

\footnotetext[2]{This error represents the region which encompasses 68\% of the data points from the best fit}

\begin{figure}[t] 
\includegraphics[width=0.5\textwidth]{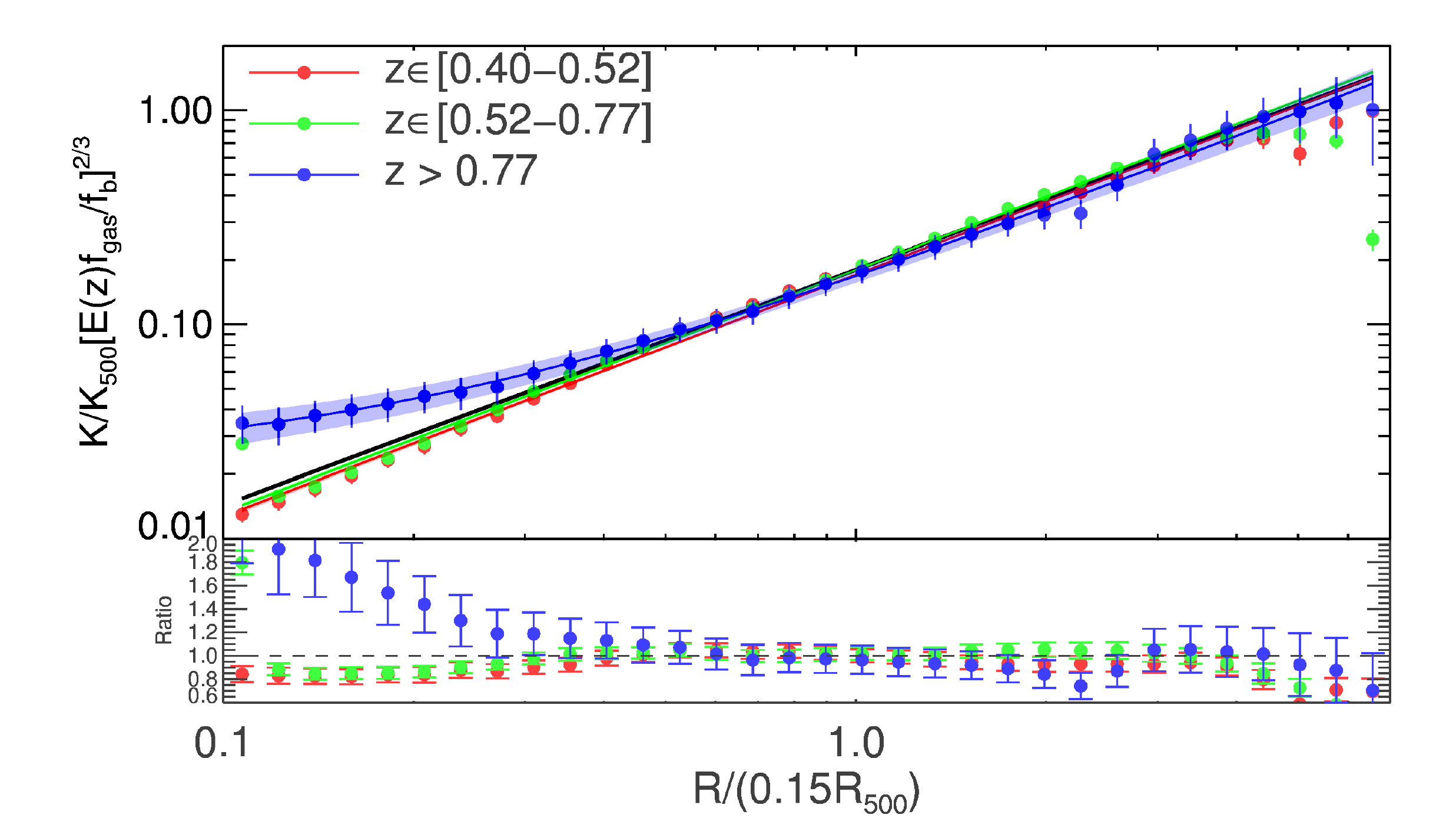}
\caption{Fits in different redshift bins showing the errors on the fits using the shaded area. 
The blue profile ($z>0.77$) is  compatible with the low redshift profiles for intermediate radii ($R>0.07R_{500}$) 
while at low radii there is a difference with the high redshift entropy profile having a flat profile.
The black line represents the non-radiative prediction \citep{voit+05}
}
\label{Figure:f_gas+K_fgas}
\end{figure}

\subsection{Rescaling using $K_{500}$}

The profile of a thermodynamical quantity, including entropy, should have a unique shape for galaxy clusters, after adequate rescaling \citep{voit05}.
Non-radiative simulations \citep[see][]{voit+05} have shown that we should rescale using quantities defined with respect to the critical density in order to achieve this. 
As it was stated above, we use an overdensity of 500, and rescale the entropy profile using $K_{500}$, defined in Eq. \ref{eq:K500}. 
We can observe the effect of the scaling in the third panel of Fig.~\ref{fig:rescaling}.
The profiles we get are less scattered than the raw data, even though the dispersion about the mean is still quite high, about one order of magnitude. 
Nevertheless we observe that above $0.15 R_{500}$ most of the clusters have a self similar behaviour. 
It is due to the fact that non gravitational processes are less relevant in the outskirts of galaxy clusters \citep{voit05}.

As described in Section 4, we stack radially the data and we fit using Eq. (\ref{eq:abc}). We show the regrouped data and the fitting results with their errors bars in Fig.~\ref{Figure:K500_errorevolution} and in top part of Table~\ref{Table:fit_fb}. 

As we can observe from the ratio between the data and the predicted profile (bottom panel of Fig. \ref{Figure:K500_errorevolution}), self-similarity is present below $0.6 R_{500}$ for the two low redshift bins while it is reached 
only between $0.15R_{500}$ and $0.6R_{500}$ at high redshift ($z > 0.77$).
Moreover the high redshift stacked profile is slightly flatter than the others, with a slope of $1.0 \pm 0.1$ and with its best fit requiring a constant term different from 0 in order to reproduce the data.

In the top part of Table~\ref{Table:fit_fb} we show the best fit dividing the sample in CC and NCC. This parameters refers to Fig.~\ref{fig:KCCNCC}. As we can see both from the best fit and the plot, CC clusters are compatible with non-radiative prediction \citep{voit+05}, with both slope and normalization. On the other hand the NCC subset of clusters has a rather flat entropy profile.

\subsection{Rescaling using the gas fraction}

\begin{figure*}[ht!]
\includegraphics[width=\textwidth]{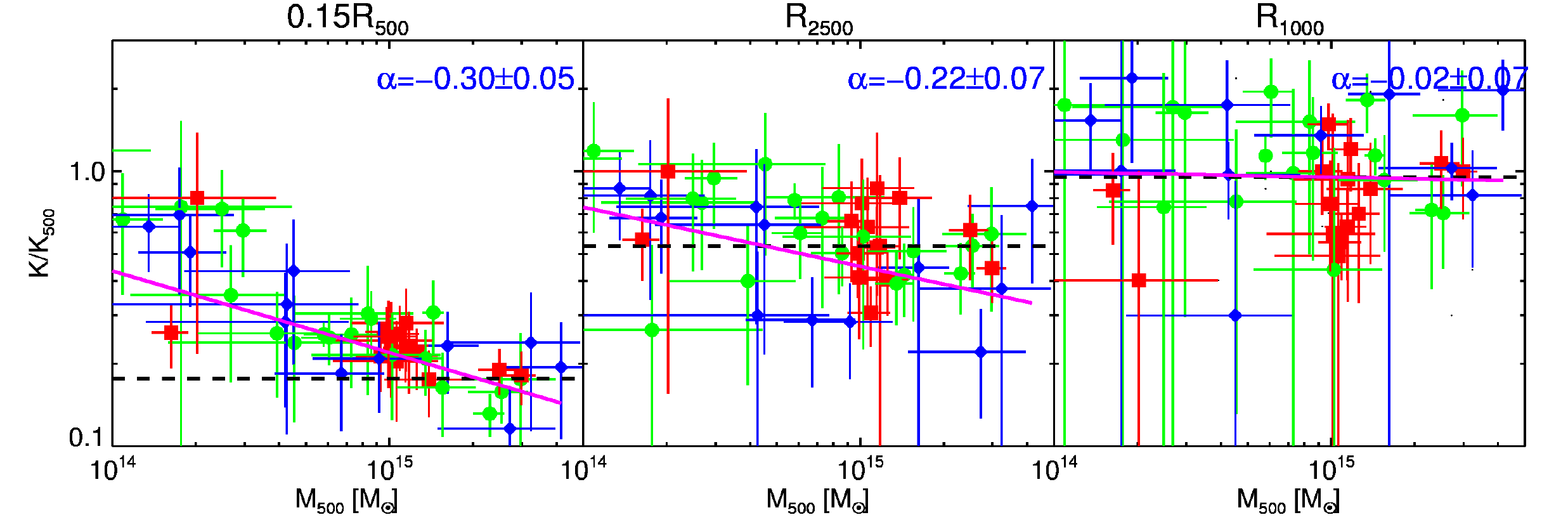}
\caption{Dimensionless entropy $K/K_{500}$  as a function of mass $M_{500}$ at different radii. Here the radii of  $0.15R_{500}$, $R_{2500}$ and $R_{1000}$ have been used. The black dashed curve is the expectation from the self-similar model, and the magenta line is the best fit using a power law with index $\alpha$  on all the data points. Red, green and blue points indicate data points from $z \in [0.4-0.52]$, $z \in [0.52-0.77]$ and $z > 0.77$ respectively }
\label{fig:variK}
\end{figure*}

The entropy distribution depends on baryon fraction with a mass (or equivalently, temperature) dependence. Consequently entropy has both a radial and global dependence on the gas fraction \citep{pratt+10}. Correcting by this effect, data become compatible with the non-radiative prediction \citep{voit+05} and the dispersion drops dramatically (last panel of Fig.~\ref{fig:rescaling}).

A practical way to quantify the deviation from the self-similar prediction is shown in Fig.~\ref{fig:variK}, where we show the behaviour of the dimensionless entropy profiles at some specific radii ($0.15R_{500}$, $R_{2500}$ and $R_{1000}$)\footnote{Radii as big as $R_{500}$ are extrapolated and would not yield robust results and are therefore not shown.}, by interpolating the surrounding data points, with respect to the mass. We note that when the mass decreases the deviation from the self-similar prediction increases. By modelling this dependance with a simple power law, we obtain a slope value which becomes smaller (in modulus) at larger radii. In particular, we point out that at the highest radius considered ($R_{1000}$), the profile is compatible with a flat one, even though the influence of just 2 or 3 points makes the best fit slope slightly negative.
 
We renormalized the entropy profiles multiplying them by gas fraction profiles ($K \rightarrow K \times (E(z)f_{gas}(R)/f_b)^{2/3}$). 
The resulting profiles are visible in the last panel of Fig.~\ref{fig:rescaling}. The self similarity of the entropy profile is now finally clear in our dataset of clusters at different redshift.
At $0.15R_{500}$, for instance, the scatter is reduced by a factor $\sim$3 when the rescaling by the gas fraction is applied.

We then stack out profiles and we show the resulting entropy radial distribution in Fig.~\ref{Figure:f_gas+K_fgas}. We notice that the stacked profiles are compatible with the non-radiative prediction \citep{voit+05} in the radial range [$0.05R_{500}$--$0.7R_{500}$], with the two low redshift stacked profile pushing this compatibility down to the lower limit of our analysis. Moreover at large radii, $R > 0.7 R_{500}$, all the stacked profiles are slightly below the prediction.

We fit the stacked entropy profiles with a power law plus a constant and we show the results in Fig.~\ref{Figure:f_gas+K_fgas} and bottom part of  Table~\ref{Table:fit_fb}. We notice that the goodness of the fit has improved, with respect to the rescaling without gas fraction, and the parameters we get are closer to the non-radiative prediction.

The slope values we obtain are slightly larger than 1.1, indicating profiles steeper than the simulated one, in agreement with several recent work \citep{voit+05,mcdonald+13,morandi+07,cavagnolo+09}. The situation is different in the case of $z > 0.77$, where the value of the central entropy is significantly different from zero.
This indicates a high average central entropy for clusters at high redshifts.

In the bottom part of Table~\ref{Table:fit_fb} we show the best fit dividing the sample in CC and NCC. This parameters refers to Fig.~\ref{fig:KfCCNCC}, the plot analogous to Fig.~\ref{Figure:f_gas+K_fgas}. Similarly to what happens to the split in redshift, entropy profile of CC gets steeper than the prediction. The NCC are still flatter than the prediction and need a central entropy different from zero in order to fit the data.

\begin{table}
\begin{center}
\resizebox{0.5\textwidth}{!} {
\begin{tabular}{ c c c c c }

\hline
subset & $K^{'}_{0.15}$ & $\alpha'$ & $K^{'}_{0}$ & $\chi_{red}^2$ \\
\hline
$z \in [0.4 , 0.52]$ & $0.188 \pm 0.003$ & $1.04 \pm 0.01$ & -- & $2.34$ \\

$z \in [0.52 , 0.77]$ & $0.189 \pm 0.002$ & $1.04 \pm 0.01$ & -- & $2.0$\\

$z > 0.77$ & $0.16 \pm 0.02$ & $1.0 \pm 0.1$ & $0.04 \pm 0.01$ & $0.14$\\

CC & $0.192 \pm 0.003$ & $1.12 \pm 0.02$ & -- & 0.54\\

NCC & $0.217 \pm 0.003$ & $0.77 \pm 0.01$ & -- & 1.86\\
\\
\hline
subset & $K^{''}_{0.15}$ & $\alpha''$ & $K^{''}_{0}$ & $\chi_{red}^2$ \\
\hline
$z \in [0.4 , 0.52]$ & $0.171 \pm 0.002$ & $1.13 \pm 0.01$ & -- & $1.26$ \\

$z \in [0.52 , 0.77]$ & $0.179 \pm 0.002$ & $1.13 \pm 0.01$ & -- & $3.68$\\

$z > 0.77$ & $0.15 \pm 0.01$ & $1.16 \pm 0.08$ & $0.022 \pm 0.005$ & $0.18$\\

CC & $0.173 \pm 0.002$ & $1.18 \pm 0.01$ & -- & $0.46$\\

NCC & $0.184 \pm 0.005$ & $0.97 \pm 0.03$ & $0.006 \pm 0.003$ & $1.54$\\

\hline
\end{tabular}
}
\end{center}
\caption{Best-fitting values, and relative errors, of the parameters of the models on entropy rescaled by $K_{500}$ (Top) (Eq.~\ref{eq:abc} ) and on entropy rescaled by both $K_{500}$ and gas fraction (Bottom) (Eq.~\ref{eq:abcd}). 
In the bin $z>0.77$, we have a value for the central entropy which indicates the presence of much more NCC systems at high redshift. 
The exponent $\alpha$ and the term $K_{100}$ are compatible with the prediction of \cite{voit+05}. 
We also see that the goodness of the fit improves a lot when we correct by the gas fraction.}
\label{Table:fit_fb}
\end{table}

\section{Discussion}

In the recent past, the dichotomy cool core clusters with a steep density profile and a drop of the temperature in the center, and non cool core clusters with a rather flat density profile and a flat temperature profile in the center, has been studied in numerous works.
The particular shape of the density profile reflects in the behaviour of the entropy profile, where CC have a low entropy floor in the center while NCC have a higher one (see Fig.~\ref{fig:KCCNCC} and \ref{fig:KfCCNCC}).

\subsection{Self-similarity}

Non-radiative simulations predict that the thermodynamic properties of clusters of galaxies should be self-similar once rescaled to specific physical quantities. In the previous section we have shown that using a proper rescaling we reach self-similarity (see Fig.~\ref{fig:rescaling}). 
Self-similarity is observed for all the redshift ranges.
However, only a proper rescaling using the gas fraction makes the agreement with the prediction \citep{voit+05} within 20\% above $0.05R_{500}$.
This is shown in the bottom panel of Fig.~\ref{Figure:K500_errorevolution} and  \ref{Figure:f_gas+K_fgas}, where the ratios between data rescaled as indicated in Eq.~\ref{eq:abc} and \ref{eq:abcd}, respectively, and the non-radiative prediction \citep{voit+05} is shown.
We observe that, only in the high redshift bin the rescaling by the gas fraction is needed to recover the self similar behaviour, reached between 0.05 and 0.7 $R_{500}$ within 20\% from the theoretical value. 

The deviations present in the inner part of the profiles may be interpreted as some form of residual energy (\cite{morandi+07}), which may be due to some non-gravitational physics processes.

Even though the agreement with simulations is remarkable, we get slightly steeper entropy profiles: $1.13 \pm 0.01$ for the low redshift bins, while we get a flatter profile in the high redshift bin, with power law index compatible with the prediction, $1.16 \pm 0.08$, but with a non zero central entropy. 
 
The agreement with the results of \cite{cavagnolo+09} is excellent in all three redshift bins since we find all slopes between 1.1 and 1.2. For the high redshift bins adding a constant to reproduce the inner part of the profile is required in order to make our results on the slope compatible with both non-radiative predictions \citep{voit+05} and observations \citep{cavagnolo+09}. Otherwise a simple power law would yield a flat profile, with slope $0.90 \pm 0.03$. 
 
The excess with respect to \cite{voit+05} self-similar prediction is present at low radii where most of our data are above the prediction. 
This extra entropy is more pronounced in low mass systems, as shown in Fig.~\ref{fig:variK} and consistent with  
the results obtained by \cite{pratt+10}.

\subsection{Angular resolution effect}

\begin{figure}[t]
\includegraphics[width=0.5\textwidth]{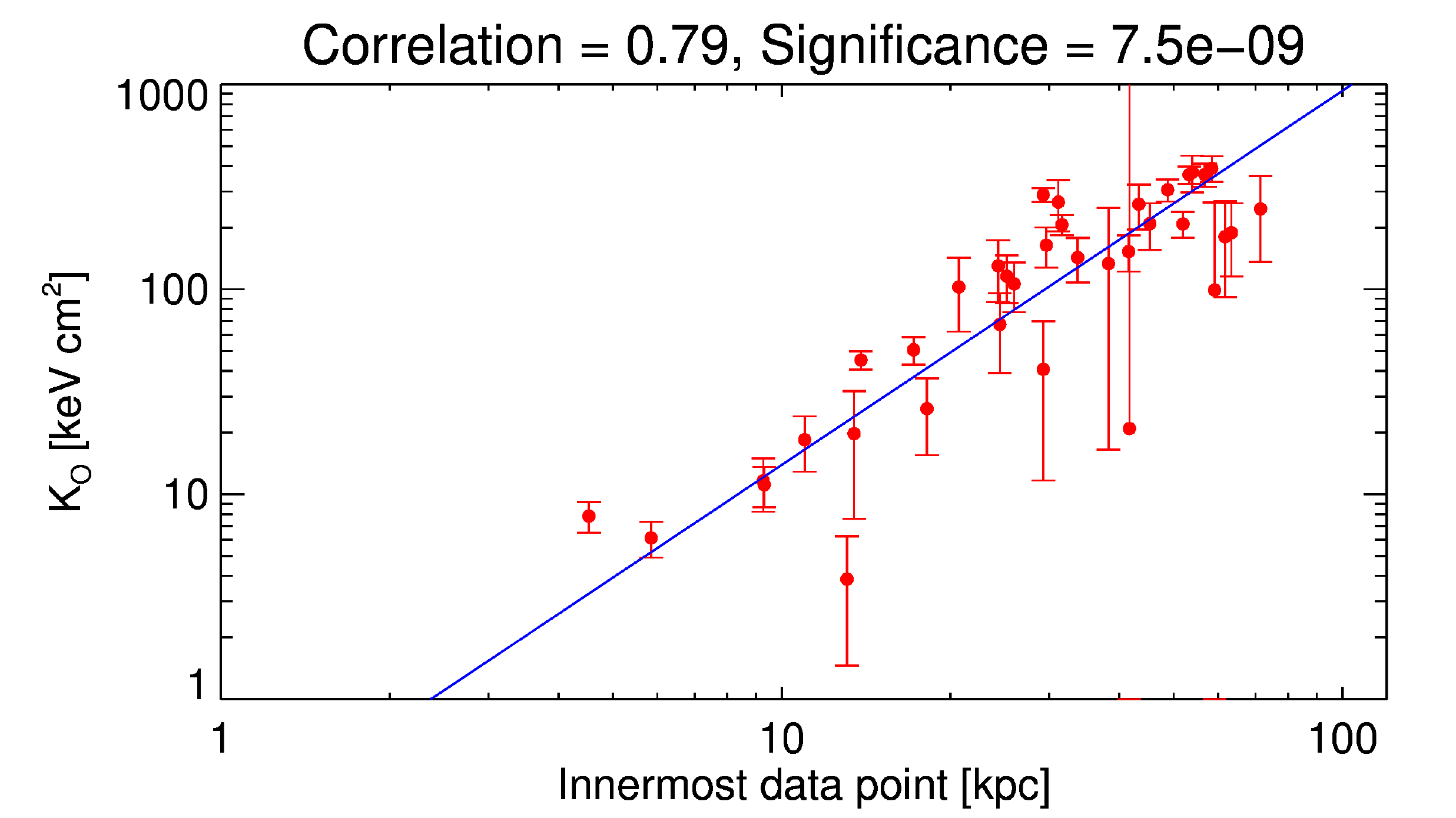}
\includegraphics[width=0.5\textwidth]{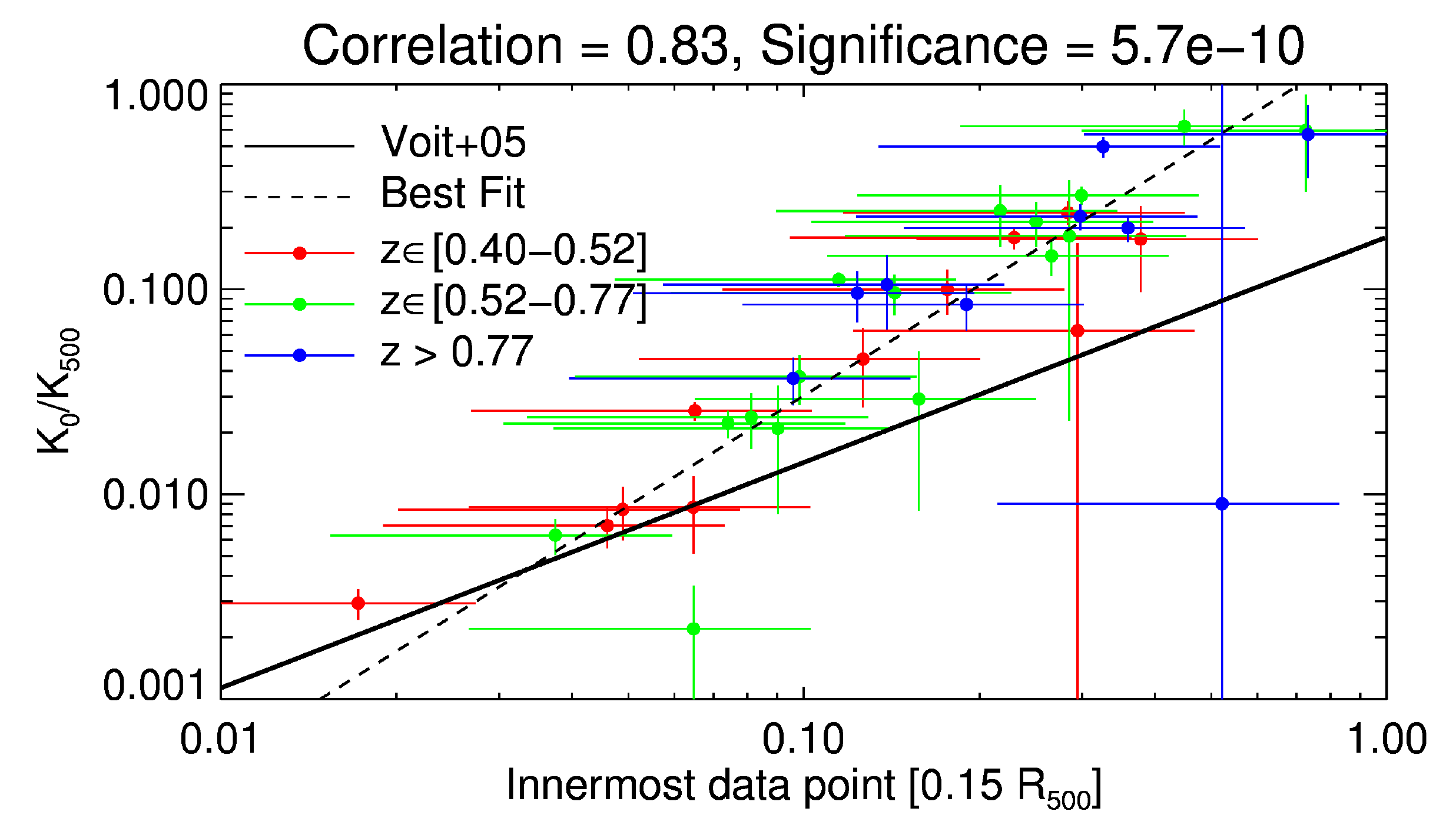}
\caption{(Top) Central entropy versus the innermost data point. A clear postiive correlation is measured. 
(Bottom) Central entropy versus innermost data point rescaled, an even tighter correlation is present.
}\label{fig:k0bin}
\end{figure}

In a thermalized system, low entropy gas sinks in the center while high entropy gas floats out in the outskirts, producing an entropy profile that increases monotonously with the radius.
The net effect is that the larger is  the central bin considered for the analysis, the higher will be the measured value of the entropy. 
In the top panel of Fig.~\ref{fig:k0bin}, we plot the value of $K_0$ for each cluster versus the radius of the innermost data point and measure 
an evident positive correlation (Pearson's rank correlation of 0.76, corresponding to a significance of $8.6 \times 10^{-8}$ of its deviation from zero that is 
associated to the case with no correlation).
This correlation becomes even more significative if we consider the same points rescaled by the halos' properties. 
This is shown in the bottom panel of Fig.~\ref{fig:k0bin}, where we measure a very significant Pearson's rank correlation of 0.83.
A similar result was shown by \cite{panagoulia+14}: the smaller the innermost radial bin, the smaller the central entropy we measure.

However, as we show in Fig.~\ref{fig:k0bin}, the correlation between the rescaled central entropy and the innermost radial bin 
 shows a slope value of $1.78 \pm 0.04$, which is about 8$\sigma$ away from the expected value, and therefore does not reproduce the predicted radial dependence from \cite{voit+05} (see Equation \ref{eq:voit}),
suggesting that the flattening has a different origin form the lack of spatial resolution.

\subsection{Note on the sample completeness}

In \cite{amodeo+16}, an extended study on the completeness of the sample analysed here is performed. In their Fig.~13, the completeness functions are compared to the distribution of the objects in the $M_{200}-z$ plane, showing that the applied selection criteria effectively selected the very massive high end of the cluster halo function in the investigated redshift range.

The observed trend to detect lower masses at higher redshifts is intrinsic to the halo mass function predicted for the hierarchical structure formation in cold dark matter dominated universe.

Therefore, we conclude that, within the limits of our sample selection and statistics, the observed redshift and mass dependences are a reasonable representation of the behaviour of the X-ray luminous cluster population in the high mass end between $z=0.4$ and $z=1.2$.

\subsection{Bimodality}

\cite{cavagnolo+09} have shown that the distribution of the values of the cluster central entropy reflects the dichotomy between CC and NCC clusters, finding two distinct populations peaking at 15 keV cm$^2$ and 150 keV cm$^2$. This bimodality has not been confirmed in later work
\citep[e.g.][]{santos+10,pratt+10}.

In Section 3.2 we have shown that using the BIC there are no statistical differences between a unimodal and a bimodal distribution. 
We consider if this lack of statistical evidence is due to the poor statistic of the cluster sample.  We build a bootstrap analysis of our distribution by selecting 10,000 samples of our objects (also allowing for repetitions). This approach permits us to determine whether a random sample extracted from our data will show an evident bimodal distribution. The results of this analysis are shown in Fig.~\ref{fig:BIC}. We observe a median value of $2.0_{-5.6}^{+3.2}$ and conclude that this BIC's distribution is compatible with no differences between a unimodal and a bimodal distribution.

In Fig.~\ref{fig:bimodality_evolution} the effect of redshift on the distribution of the central entropy is shown.
As the number of clusters in each redshift bin is too small, we are not able to accurately prove any redshift evolution of the central entropy, however we have indications that an evolution may be present. 

At high redshift we have an important peak at entropy higher than 100 keV cm$^2$, which represents the NCC systems, and only one cluster with central entropy lower than 10 keV cm$^2$. 

We observe that from high redshift to low redshift the peak at high entropy becomes less prominent and a larger fraction of clusters get a smaller value of the central entropy, so that in the lower redshift bins the majority of our clusters have a central entropy below 100 keV cm$^2$.

Nevertheless there is an indication of evolution from many NCC systems at high redshift toward mostly relaxed CC clusters at low redshift. 
Due to the poor statistic of the sample we are not able to prove this scenario using statistical tests.

\subsection{Evolution with redshift}

We observe, within the central region, an evolution with redshift of the entropy  ($r < 0.1 R_{500}$). 
It suggests that the entropy profiles are flatter at high redshifts in massive objects, or, alternatively, there is a non-zero central entropy at high redshift. 
This resembles an evolution in the entropy profiles. However, as we show in Fig.~\ref{fig:bimodality_evolution}, the fraction of NCC clusters is much higher at high redshifts than at low redshifts. 
This would imply that at $z > 0.77$ the clusters in our dataset are not able to develop a cool core like clusters at low redshift do, meaning that cool cores were less common in the past, flattening the entropy profiles and creating non zero value for the central entropy. Several studies \citep[e.g.][]{vikhlinin+07,santos+08,santos+10,mcdonald+11}, that have investigated the evolution of the cool-coreness of clusters, support this scenario, in particular on the lower relative abundance of the strongest CC at high redshift.
In other works \citep[e.g.][]{mcdonald+13,mcdonald+17}, the ``cuspiness'' of the gas density is shown to decrease with increasing redshift, as consequence of a non-evolving core embedded in an ambient ICM which evolves self-similarly.

If we define the CC clusters as the one with central entropy lower than 100 keV cm$^2$ then from Fig.~\ref{fig:bimodality_evolution} we derive that in the low redshift bins 67\% of clusters are CC, percentage which reduces to 50\% and 40\% in the intermediate and high redshift bins respectively.


This result slightly deviates, but not in a significant way, from what has been presented in previous studies \citep[e.g.][]{vikhlinin+07,mcdonald+13}.
For example, \citep[][see their fig.12]{mcdonald+13} obtain that CCs represent 40\% (10--40\%) of the cluster population at low (high) redshift. Their sample is SZ selected and is, therefore, less biased toward CC clusters with respect to an X-ray selected sample.

This result is consistent with the hierarchical scenario of the growth of structure, given that at high redshift ($z \sim 1$) clusters were in the middle of their formation history and cool cores could have easily been destroyed by one of the many merger events, or not even built, if they did not have enough time to relax in the center. 
Moreover, it is remarkable that even in our sample, selected in order to have specific X-ray properties and thus prone to include X-ray bright centrally-peaked cool core objects, we consider a relative larger fraction of NCC at higher redshift.

\begin{figure}[t]
\vbox{
\includegraphics[width=0.5\textwidth]{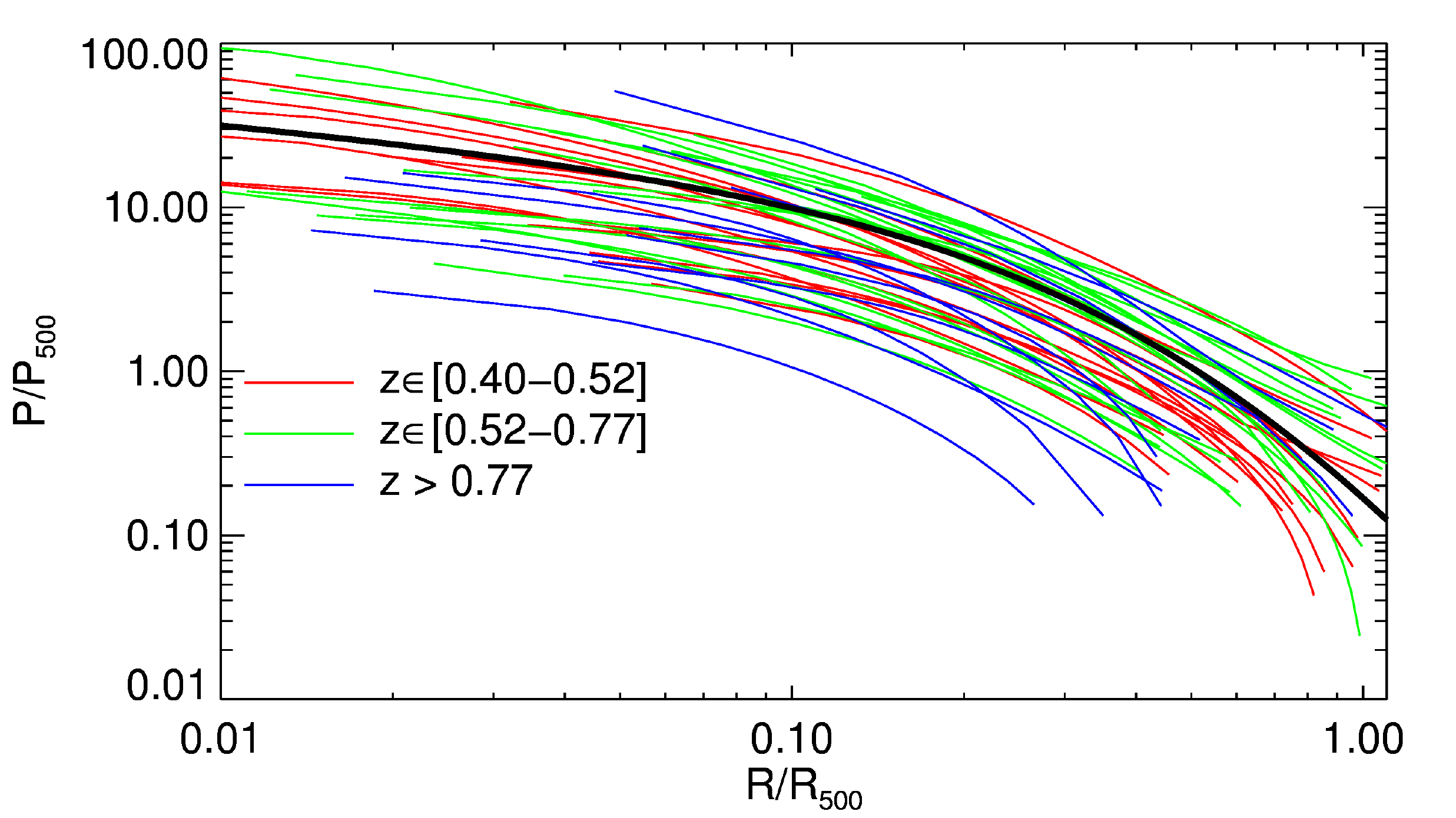}
\includegraphics[width=0.5\textwidth]{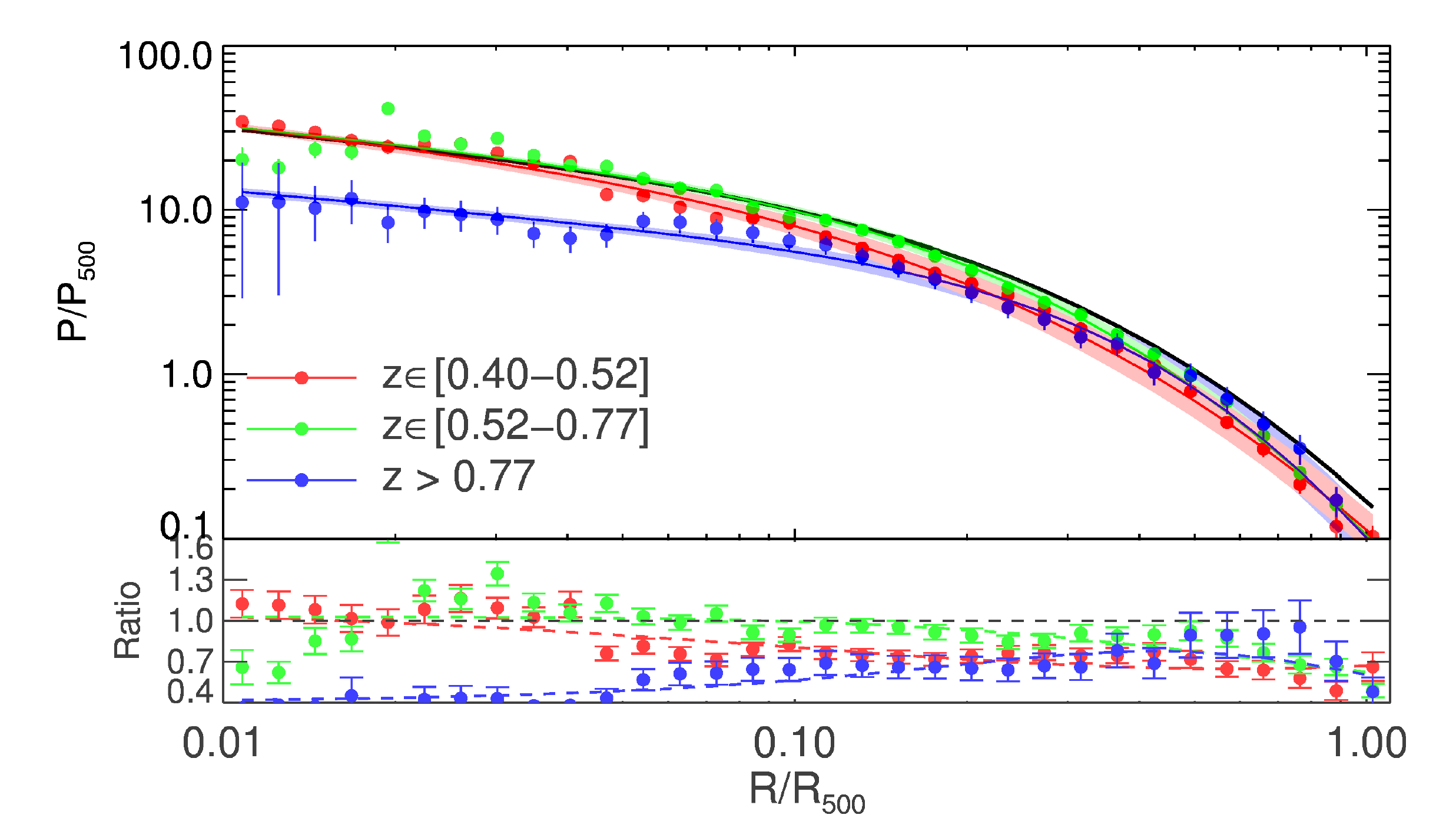}
}
\caption{(Top) Pressure profiles for all our clusters rescaled using an overdensity of 500 (color coded with respect 
to the redshift bin to which each curve belongs), compared with the best fit results in \cite{arnaud+10} (black solid line).
(Bottom) Stacked pressure profiles (in logarithmic space) compared with the best fit of \cite{arnaud+10}; the panel at the bottom shows the ratio with respect to the ``universal'' pressure profile.
}
\label{fig:pressure}
\end{figure}

\subsection{Pressure}
\begin{table*}[t]
\begin{center}
\resizebox{0.77\textwidth}{!} {
\begin{tabular}{ c c c c c c c }
\hline
subset & $P_0$ & $c_{500}$ & $\gamma$ & $\alpha$ & $\beta$ & $\chi^2_{red}$  \\
\hline
$z \in [0.4 , 0.52]$ & $9.1 \pm 0.3$ & $1.08 \pm 0.08$ & 0.308 & $0.87 \pm 0.04$ & 5.49 & 2.1\\

$z \in [0.52 , 0.77]$ & $9.2 \pm 0.2$ & $1.48 \pm 0.05$ & 0.308  & $1.11 \pm 0.04$ & 5.49 & 4.6 \\

$z > 0.77$ & $3.6 \pm 0.2$ & $1.5 \pm 0.1$ & 0.308 & $1.59 \pm 0.14$ & 5.49 & 0.8\\

CC &  $11.3 \pm 0.3$ & $1.30 \pm 0.07$ & 0.308 & $0.92 \pm 0.03$ & 5.49 & 2.1\\

NCC & $5.46 \pm 0.14$ & $1.57 \pm 0.05$ & 0.308 & $1.43 \pm 0.06$ & 5.49 & 1.9\\

\\
Universal & 8.403 & 1.177 & 0.308 & 1.0510 & 5.49 & --
\end{tabular}
}
\end{center}
\caption{Best-fitting values, and relative errors, for the pressure profiles modelled with the generalized NFW in equation~\ref{eq:arnaud}. 
$\beta$ and $\gamma$ are frozen to their ``universal'' values in order to break some degeneracy and facilitate the comparison between the results.
The first column indicates the subset of our sample chosen to fit the data. We observe that the results on $c_{500}$ and $\alpha$ of NCC subsample and the result of high redshift subeset are compatible within 1$\sigma$
The goodness of the fit is indicated in the last column on the right. In general the value of the parameters are quite close to the ``universal'' results of \cite{arnaud+10} considering that they are highly degenerate.
We point out that the parameter $P_0$  for the high redshift bin is significantly smaller than its value on the other bins and the ``universal'' value.
} 
\label{Table:fit_nagai}
\end{table*}	
\begin{figure*}
\hbox{
\includegraphics[width=0.5\textwidth]{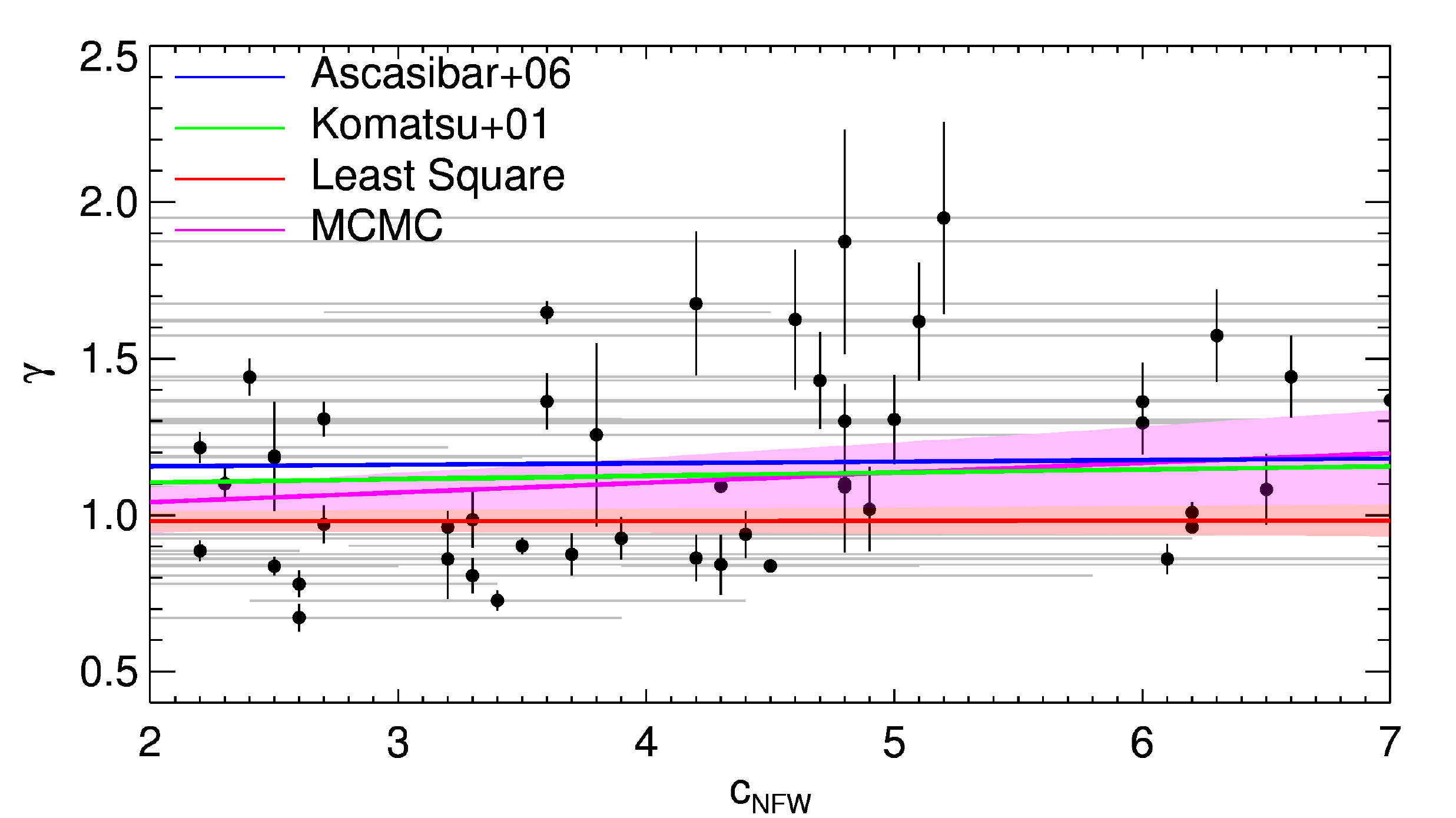}
\includegraphics[width=0.5\textwidth]{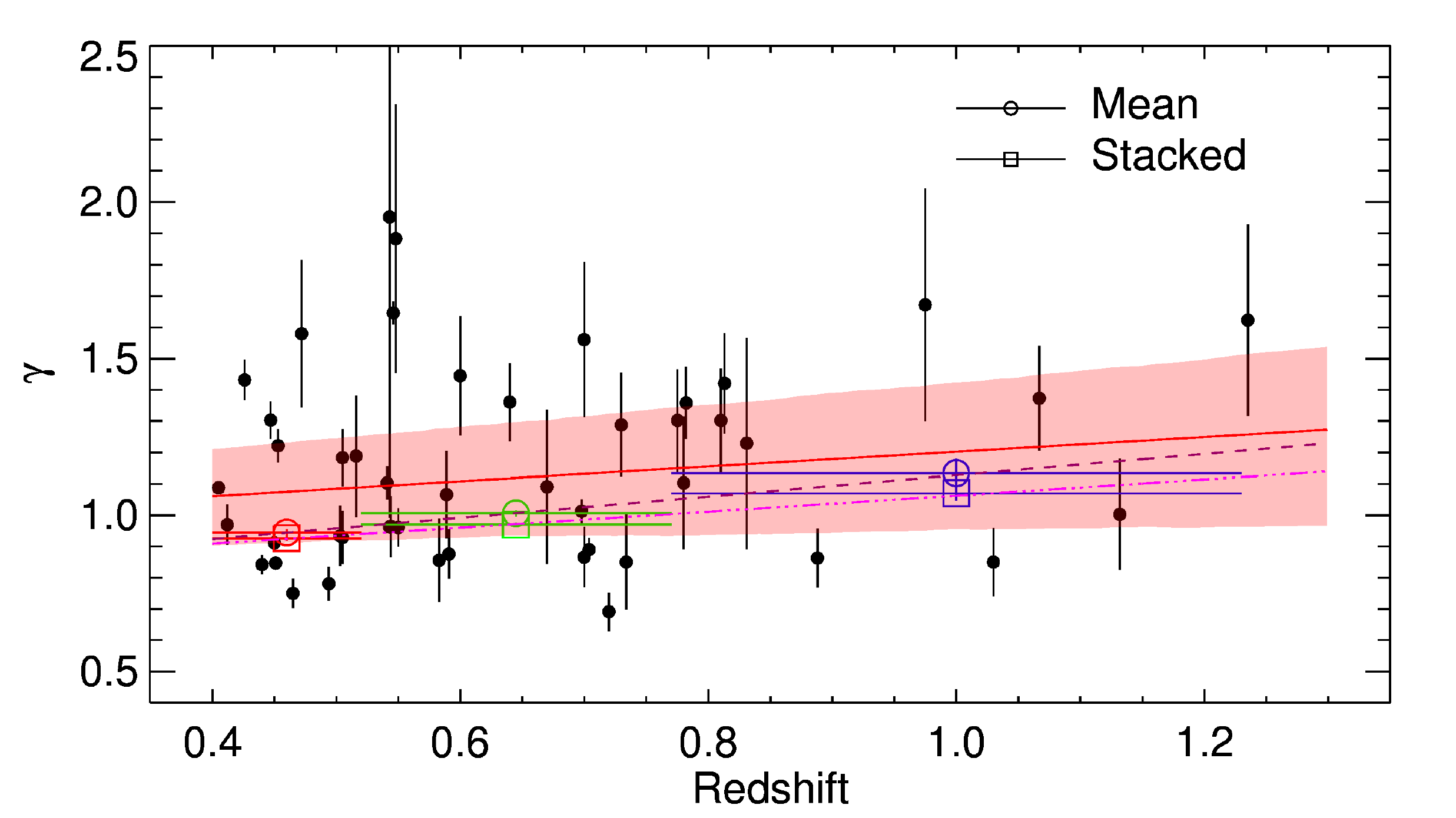}
}
\caption{(Left) Polytropic index for the objects in our sample versus concentration: $\gamma$ increases with larger values of $c_{\text{NFW}}$. 
 Black points are the observed data points, green and blue line are the results of \cite{komatsu+01} and \cite{ascasibar+06} respectively, the red line with pink shaded area is the best fit using classical least square minimization with 1$\sigma$ confidence region, and the magenta line with lily shaded area is the best fit usinc an MCMC algorithm.  
(Right) Polytropic index for all clusters as function of redshift. Data are better fitted by a linear relation: the effective polytropic index grows with redshift. 
Black points are the observed data points, the red line with pink shaded area are the the best fit with 1$\sigma$ dispersion using MCMC algorithm. The empty circles are the value of adiabatic index using the mean result for the clusters in each bin, while the empty squares are the polytropic index of the stacked profiles in each redshift bin, with the brown dashed line and the dash-dotted magenta line being the best linear fit on the ``mean'' and ``stacked'' adiabatic index.
The parameters describin these lines are listed in Table~\ref{Table:poly}.
}
\label{fig:gamma}
\end{figure*}
For the same dataset, we study the behaviour of the electronic pressure profile $P(r) = n_{\rm e}(r) \, T(r)$. 

A generalized NFW profile, as introduced by \cite{nagai+07} (see Equation~\ref{eq:arnaud}), has been widely used to study the radial rescaled pressure profile.
The best fitting parameters $[P_0, c_{500}, \gamma, \alpha, \beta] = [8.403, 1.177, 0.3081, 1.0510, 5.49]$ obtained from \cite{arnaud+10} 
represent the so called ``universal pressure profile'' for galaxy clusters.

Pressure is the quantity less affected by the thermal history of the cluster \citep{arnaud+10}. 
Indeed \cite{mcdonald+14} found no significant evolution of the pressure profile in the analysis of SPT SZ-selected clusters, just a mild flattening of the profile below $0.1 R_{500}$. 
\cite{battaglia+12}, however, suggested from the analysis of cosmological hydrodynamical simulations that 
a significant evolution of the pressure profile should occur beyond $z = 0.7$, and only outside $R_{500}$,
as consequence of the increasing non-thermal support toward the outskirts of galaxy clusters. 

In the functional form shown in Equation \ref{eq:arnaud}, we fix the parameters $\beta$ and $\gamma$ to the fiducial value found in the work of \cite{arnaud+10}. These parameters are degenerate, and therefore fixing at least one of the slopes is advised in order to have tighter parameters distribution and  better  comparison \citep{arnaud+10}.

The pressure profiles are plotted in the top panel of Fig.~\ref{fig:pressure}, together with the curve of the universal profile, in order to make a comparison. 
All the profiles show a very similar shape, surrounding the ``universal'' one from both sides with an apparent discrepancy only in the normalization of the profiles.
At each radius the scatter is of about one order of magnitude.

We applied the same procedure described in Section~\ref{ana} and used in the analysis of the entropy profiles:
we interpolate over the same radial grid, extrapolate up to $R_{500}$ using the best-fit model, and stack these pressure profiles.
We obtain that the stacked curves in the two low redshift bins are compatible with the ``universal'' pressure profile at radii below $0.1 R_{500}$, while above it the data points are slightly below it. Nevertheless the best fit on these stacked data points comprehend the \cite{arnaud+10} result at all radii within the error bars.

On the other hand the high redshift scaled profile for the inner part of these profile we observe a distinctive flattening below $0.1 R_{500}$, with values which are about 30\% of the ``universal'' pressure profile. Above this radius the data points are very close to the other redshift bins, meaning slightly below the \cite{arnaud+10} result.

Pressure is indeed a thermodynamic property that is only very little affected by the thermodynamic history of clusters. The observed flattening at high redshift at low radii was also observed in the work of \cite{mcdonald+14} and can be easily explained by the little presence of CC clusters at high redshift. 
 
In Table~\ref{Table:fit_nagai} we show the best fit of Equation \ref{eq:arnaud}, where we have kept $\beta$ and $\gamma$ fixed to the ``universal'' pressure profile best fit. We obtain a quite good fit in all the three redshift  bins; only the normalization of this functional in the high redshift bin shows a distinct discrepancy with the results of \cite{arnaud+10}, greater than 5$\sigma$. In fact from the bottom panel of Fig.~\ref{fig:pressure} is evident that the low redshift bins points are almost on top of the ``universal'' profile, while the high redshift ones are compatible only above $0.1 R_{500}$, and below this threshold the discrepancy grows to be factor 3 at $0.01 R_{500}$.
For completeness in the same table we show the fitting done for the subsets of CC and NCC, referring to the data in Fig.~\ref{fig:PCCNCC}. We observe that NCC clusters best fit results resembles the high-z subsample, while CC clusters resemble low-z objects.

\subsection{Polytropic index}

The polytropic index $\gamma$, equal to the ratio of specific heats $C_P/C_V$ for an ideal gas, is a common proxy when evaluating the physical state and the thermal distribution of the gas. 
It is defined as
\begin{equation}
P_e = cost \cdot n_e^\gamma,
\end{equation}
with values of $\gamma$ expected to be in the range [1.1, 1.2] \citep{komatsu+01,ascasibar+06,shi+16}, i.e. between 1, the value describing an isothermal gas, and 5/3, the value of an isoentropic gas, when the gas is well mixed and the gas entropy per atom is constant.

Studying the evolution of the polytropic index with redshift and its relation with the concentration $c$ of the dark matter distribution can provide a more consistent picture on 
the processes that regulate the hierarchical structure formation. A correlation between $\gamma$ and $c$ is expected if the radial structure of the ICM and of the host halo 
depend on the halo mass.
\cite{komatsu+01} require a linear relation between concentration and polytropic index by assuming in the theoretical work that the gas traces the dark matter distribution outside the core.
\cite{ascasibar+06} have shown that $c$ and $\gamma$ conspire to produce the observed  scaling relations, matching the self-similar slope at many overdensities. 

We estimate an effective polytropic index $\gamma$ by fitting the pressure with a power law as function of the gas density.
As a first step, $\gamma$ is calculated for each single cluster. Then, we evaluate the weighted mean in each bin.
We also calculate $\gamma$ for the stacked profiles. 

We look for correlations between the polytropic index and the dark matter concentration as recovered from the best-fit with a NFW model in \cite{amodeo+16}.
\cite{ascasibar+06} and \cite{komatsu+01} have shown that between concentration and polytropic index there is a linear relation with slope of 0.005 and 0.01, respectively. 
Using the Markov Chain Monte Carlo (MCMC) code \textit{emcee} \citep{emcee}, we obtain a slope much stepper that the theoretical predictions, 
although with a relative uncertainty of about 70\% (see best-fit values in Table~\ref{Table:poly}) which makes it compatible with previous results\citep{komatsu+01,ascasibar+06} within 1$\sigma$, and even compatible with 0 at 2$\sigma$.
Moreover, the intercept we get is much smaller than what has been previously calculated.

In right panel of Fig.~\ref{fig:gamma}, we show the polytropic index as a function of redshift.
We measure a positive evolution with redshift, with larger values of $\gamma$  (by more than 2$\sigma$) at higher redshift (see Table~\ref{Table:poly}).

\begin{table}
\begin{center}
\begin{tabular}{ c  c  c  }
\hline
\multicolumn{3}{c}{\textbf{Polytropic index \boldmath$\gamma$}} \\
subset & Mean & Stacked  \\
\hline
$z \in [0.4 , 0.52]$ & $0.935 \pm 0.008$ & $0.942 \pm 0.008$  \\

$z \in [0.52 , 0.77]$ & $0.996 \pm 0.009$ & $0.906 \pm 0.006$ \\

$z > 0.77$ & $1.076 \pm 0.031$ & $1.041 \pm 0.018$  \\
\hline 
\\
\multicolumn{3}{c}{\textbf{ \boldmath$\gamma$ = m $\cdot$ c$_{\text{NFW}}$ + q}} \\
Method & m & q  \\
\hline
Chi-squared & $0.0004 \pm 0.0057$ & $0.98 \pm 0.03$  \\

MCMC & $0.031 \pm 0.020$ & $0.98 \pm 0.09$ \\

Ascasibar+06 & $0.005 \pm 0.002$ & $1.145 \pm 0.007$  \\

Komatsu+01 & $0.01$ & $1.085$ \\
\hline 
\\
\multicolumn{3}{c}{\textbf{ \boldmath$\gamma$ = m $\cdot$ z + q}} \\
Method & m & q  \\
\hline
Chi-squared & $0.05 \pm 0.05$ & $0.94 \pm 0.03$  \\

MCMC & $0.24 \pm 0.22$ & $0.97 \pm 0.14$ \\

Mean & $0.29 \pm 0.05$ & $0.80 \pm 0.030$  \\

Stacked & $0.11 \pm 0.03$ & $0.86 \pm 0.02$ \\
\hline
\end{tabular}
\end{center}

\caption{(Top) Values and errors on the polytropic index for the three redshift bins considering the mean and stacked values. We observe evolution with a significance greater than 2$\sigma$.
(Middle) Fit of left panel of Fig.~\ref{fig:gamma} using different methods and comparing with previous theoretical work. 
(Bottom) Fit of right panel of Fig.~\ref{fig:gamma} using different methods.}
\label{Table:poly}
\end{table}

\section{Conclusions}
	
From the sample described in \cite{amodeo+16}, which contains the one of the largest collection of clusters at $z > 0.8$ homogeneously analyzed in their X-ray spectral properties, we have extracted the entropy and pressure profiles of 47 clusters observed with \chandra in a redshift range from 0.4 to 1.24. 

We observe higher values of the gas entropy in the central region at higher redshift, which we cannot be explained as an effect due to the spatial resolution.
A plausible explanation of this result is the fact that at high redshift we observe a lack of cool core clusters with respect to the low redshift sample. 

Moreover at intermediate radii, between $0.1 R_{500}$ and $0.7 R_{500}$, the self similarity is recovered when we use entropy dependence both on redshift and gas fraction and
the scatter between the profiles is reduced by a factor $\sim 3$.
The best fit of the stacked profiles is very similar to the \cite{voit+05} prediction from non radiative simulations. 

We also show that the pressure profiles flatten at high redshift at radii below $0.1 R_{500}$, with lower values, by about 50\%, than the ones observed at $z \la 0.5$. 

Overall, these results agrees with a scenario in which galaxy clusters are the last gravitationally-bound structures to form according to the hierarchical evolution. 
They start forming at $z \approx 3$, and at $z \sim 1-1.5$ they are still in the middle of their formation. 
At this epoch, cool cores could either be destroyed by merger events, or have not formed yet, reducing their relative number at earlier epoch. 
Moreover, the merging processes ongoing at high redshift would imply that objects at $z \sim 1$ are mostly unrelaxed, with a flatter entropy profile, which produces a clear excess in the inner parts and a deficit in the outskirts.
As we show in Figures~\ref{Figure:K500_errorevolution} and \ref{Figure:f_gas+K_fgas}, high-redshift clusters have indeed a rather flat stacked entropy profile, 
supporting the evidence that the floating and sinking of the gas entropy has not been completed yet.
The thermodynamical disturbed condition of the high-redshift systems is further supported by the observed flattening of thermal pressure 
in the inner part of the stacked pressure profile (see Fig.~\ref{fig:pressure}).

Moreover, we measure a slightly significant evolution of the effective polytropic index of the ICM, that we measure by estimating $d\log P_e / d \log n_e$, with the dark matter concentration and redshift, with 87\% significance (1.5$\sigma$) for concentration and 73\% (1.1$\sigma$) for redshift, 
indicating that the gas possesses a slightly larger polytropic index in systems which have a more concentrated mass distribution at higher redshift.
This result supports the observational evidence that at high redshift we recover more isentropic (i.e. more flat) entropy profiles.

\bibliographystyle{plain} 
\bibliography{draft_AA.bbl} 
\newpage

\begin{appendix}
\section{Additional information}
\begin{table}[h]
\begin{center}

\resizebox{0.5\textwidth}{!} {
\begin{tabular}{ c | c | c | c | c }
Object & z & $R_{500}$ & $f_{gas,500}$ & Reference for redshift \\
- &  - &  [kpc] &  [fraction] & - \\
\hline
MACS0159.8-084 & 0.405 & 1379 $\pm$ 111 & 0.11 $\pm$ 0.02 & \cite{kotov+06} \\
MACSJ2228.5+20 & 0.412 & 1418 $\pm$ 156 & 0.13 $\pm$ 0.03 & \cite{boringer+00} \\
MS1621.5+2640 & 0.426 & 1298 $\pm$ 127 & 0.1 $\pm$ 0.02 & \cite{stocke+91} \\
MACSJ1206.2-08 & 0.44 & 1874 $\pm$ 128 & 0.08 $\pm$ 0.01 & \cite{borgani+01} \\
MACS-J2243.3-0 & 0.447 & 1335 $\pm$ 137 & 0.16 $\pm$ 0.03 & \cite{coble+07} \\
MACS0329.7-021 & 0.45 & 1264 $\pm$ 113 & 0.12 $\pm$ 0.04 & \cite{allen+04} \\
RXJ1347.5-1145 & 0.451 & 1756 $\pm$ 134 & 0.1 $\pm$ 0.02 & \cite{schindler+95} \\
V1701+6414 & 0.453 & 707 $\pm$ 48 & 0.26 $\pm$ 0.04 & \cite{wang+14} \\
MACS1621.6+381 & 0.465 & 1349 $\pm$ 242 & 0.09 $\pm$ 0.04 & \cite{edge+03} \\
CL0522-3624 & 0.472 & 754 $\pm$ 271 & 0.11 $\pm$ 0.11 & \cite{mullis+03} \\
MACS1311.0-031 & 0.494 & 1420 $\pm$ 241 & 0.06 $\pm$ 0.02 & \cite{allen+04} \\
MACS-J2214.9-1 & 0.503 & 1275 $\pm$ 253 & 0.14 $\pm$ 0.06 & \cite{bonamonte+06} \\
MACS911.2+1746 & 0.505 & 1338 $\pm$ 140 & 0.1 $\pm$ 0.07 & \cite{ebeling+07} \\
MACSJ0257.1-23 & 0.505 & 1293 $\pm$ 260 & 0.12 $\pm$ 0.05 & \cite{ebeling+07} \\
V1525+0958 & 0.516 & 1259 $\pm$ 191 & 0.05 $\pm$ 0.01 & \cite{mullis+03} \\
MS0015.9+1609 & 0.541 & 1381 $\pm$ 128 & 0.13 $\pm$ 0.02 & \cite{stocke+91} \\
CL0848.6+4453 & 0.543 & 701 $\pm$ 394 & 0.13 $\pm$ 0.2 & \cite{takey+11} \\
MACS1423.8+240 & 0.543 & 1041 $\pm$ 16 & 0.13 $\pm$ 0.01 & \cite{ebeling+07} \\
MACSJ1149.5+22 & 0.544 & 1187 $\pm$ 141 & 0.19 $\pm$ 0.04 & \cite{ebeling+07} \\
MACSJ0717.5+3745 & 0.546 & 1409 $\pm$ 59 & 0.17 $\pm$ 0.02 & \cite{ebeling+07} \\
CL1117+1744 & 0.548 & 568 $\pm$ 106 & 0.2 $\pm$ 0.09 & \cite{wen+10} \\
MS0451.6-0305 & 0.55 & 1699 $\pm$ 229 & 0.07 $\pm$ 0.03 & \cite{gioia+94} \\
MS2053.7-0449 & 0.583 & 1239 $\pm$ 367 & 0.04 $\pm$ 0.02 & \cite{stocke+91} \\
MACS-J2129.4-0 & 0.589 & 1155 $\pm$ 228 & 0.15 $\pm$ 0.07 & \cite{ebeling+07} \\
MACS-J0647.7+7 & 0.591 & 1762 $\pm$ 333 & 0.06 $\pm$ 0.02 & \cite{ebeling+07} \\
CL1120+4318 & 0.6 & 894 $\pm$ 208 & 0.16 $\pm$ 0.08 & \cite{romer+00} \\
CLJ0542.8-4100 & 0.64 & 801 $\pm$ 68 & 0.13 $\pm$ 0.03 & \cite{depropris+07} \\
LCDCS954 & 0.67 & 567 $\pm$ 98 & 0.16 $\pm$ 0.07 & \cite{gonzalez+07} \\
MACS0744.9+392 & 0.698 & 994 $\pm$ 93 & 0.17 $\pm$ 0.04 & \cite{ebeling+07} \\
SPT-CL0001-5748 & 0.7 & 902 $\pm$ 261 & 0.12 $\pm$ 0.08 & \cite{vikhlinin+98} \\
V1221+4918 & 0.7 & 738 $\pm$ 127 & 0.15 $\pm$ 0.07 & \cite{mantz+14} \\
RCS2327.4-0204 & 0.704 & 1547 $\pm$ 112 & 0.08 $\pm$ 0.01 & \cite{gralla+11} \\
SPT-CLJ2043-5035 & 0.72 & 1348 $\pm$ 255 & 0.07 $\pm$ 0.02 & \cite{song+12} \\
ClJ1113.1-2615 & 0.73 & 747 $\pm$ 178 & 0.06 $\pm$ 0.04 & \cite{perlman+02} \\
CLJ2302.8+0844 & 0.734 & 1042 $\pm$ 207 & 0.06 $\pm$ 0.02 & \cite{perlman+02} \\
SPT-CL2337-5942 & 0.775 & 1590 $\pm$ 414 & 0.04 $\pm$ 0.02 & \cite{vanderlinde+10} \\
RCS2318+0034 & 0.78 & 1681 $\pm$ 484 & 0.02 $\pm$ 0.01 & \cite{hicks+08} \\
MS1137.5+6625 & 0.782 & 994 $\pm$ 203 & 0.06 $\pm$ 0.03 & \cite{gioia+94} \\
RXJ1350.0+6007 & 0.81 & 577 $\pm$ 71 & 0.17 $\pm$ 0.05 & \cite{holden+02} \\
RXJ1716.9+6708 & 0.813 & 646 $\pm$ 89 & 0.15 $\pm$ 0.05 & \cite{henry+97} \\
EMSS1054.5-0321 & 0.831 & 1308 $\pm$ 227 & 0.06 $\pm$ 0.02 & \cite{gioia+04} \\
CLJ1226.9+3332 & 0.888 & 1752 $\pm$ 409 & 0.04 $\pm$ 0.02 & \cite{ebeling+01} \\
XMMUJ1230+1339 & 0.975 & 792 $\pm$ 309 & 0.08 $\pm$ 0.07 & \cite{fassbender+11} \\
J1415.1+3612 & 1.03 & 772 $\pm$ 259 & 0.08 $\pm$ 0.05 & \cite{ellis+04} \\
SPT-CL0547-5345 & 1.067 & 779 $\pm$ 188 & 0.14 $\pm$ 0.08 & \cite{high+10} \\
SPT-CLJ2106-584 & 1.132 & 963 $\pm$ 254 & 0.13 $\pm$ 0.06 & \cite{foley+11} \\
RDCS1252-29 & 1.235 & 533 $\pm$ 124 & 0.13 $\pm$ 0.08 & \cite{rosati+04} \\

\end{tabular}
}
\end{center}
\caption{This table completes Table 2 in \cite{amodeo+16}, adding informations on the sizes and gas content of these clusters.}
\label{table:extra}
\end{table}
\newpage

\section{CC vs NCC}

\begin{figure}
\includegraphics[width=0.5\textwidth]{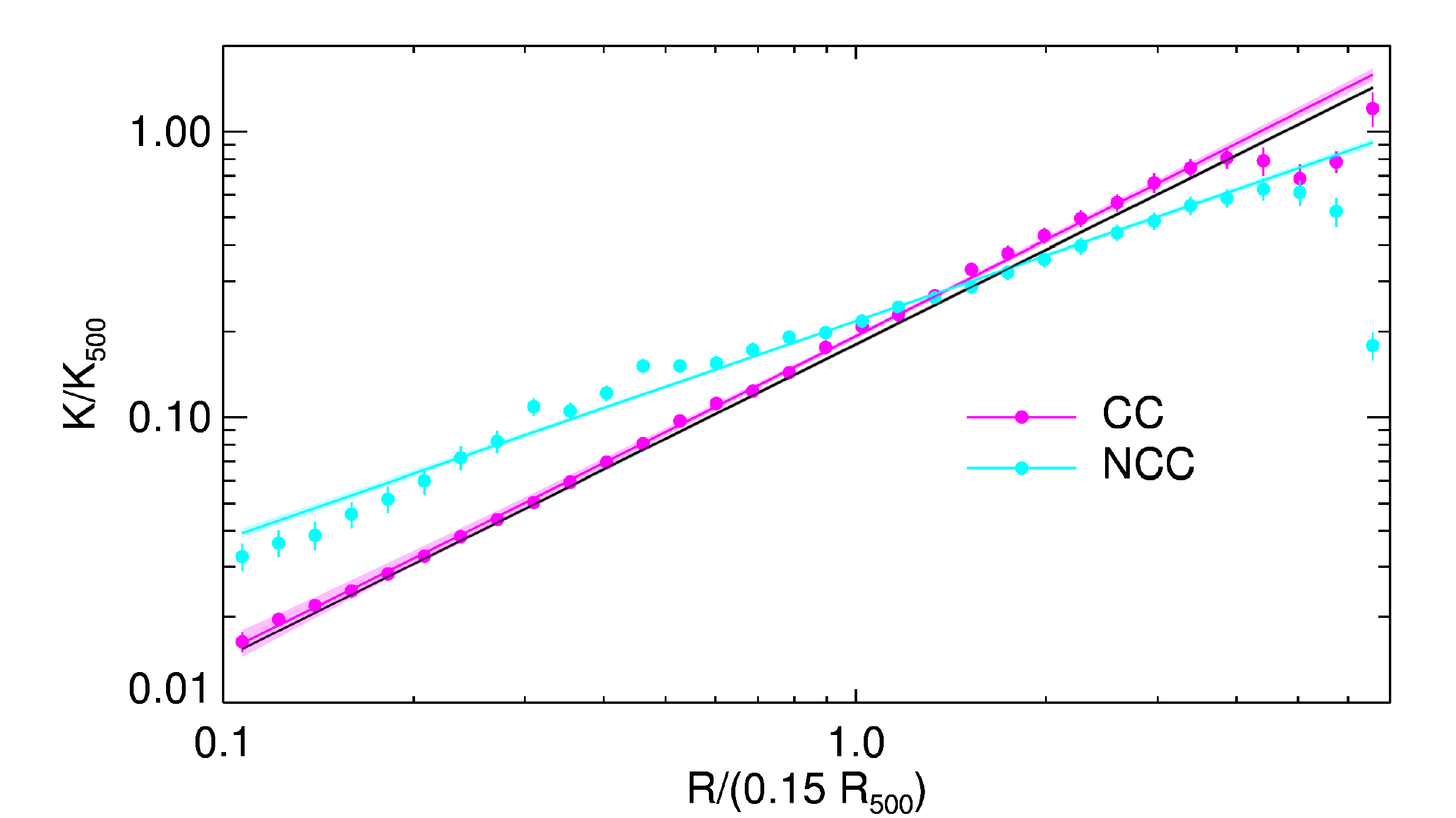}
\caption{Staked entropy profiles for the subsets of CC and NCC based on \cite{cassano+10} criteria. We observe that CC profile is almost on top of \cite{voit+05} prediction.}
\label{fig:KCCNCC}
\end{figure}

\begin{figure}
\includegraphics[width=0.5\textwidth]{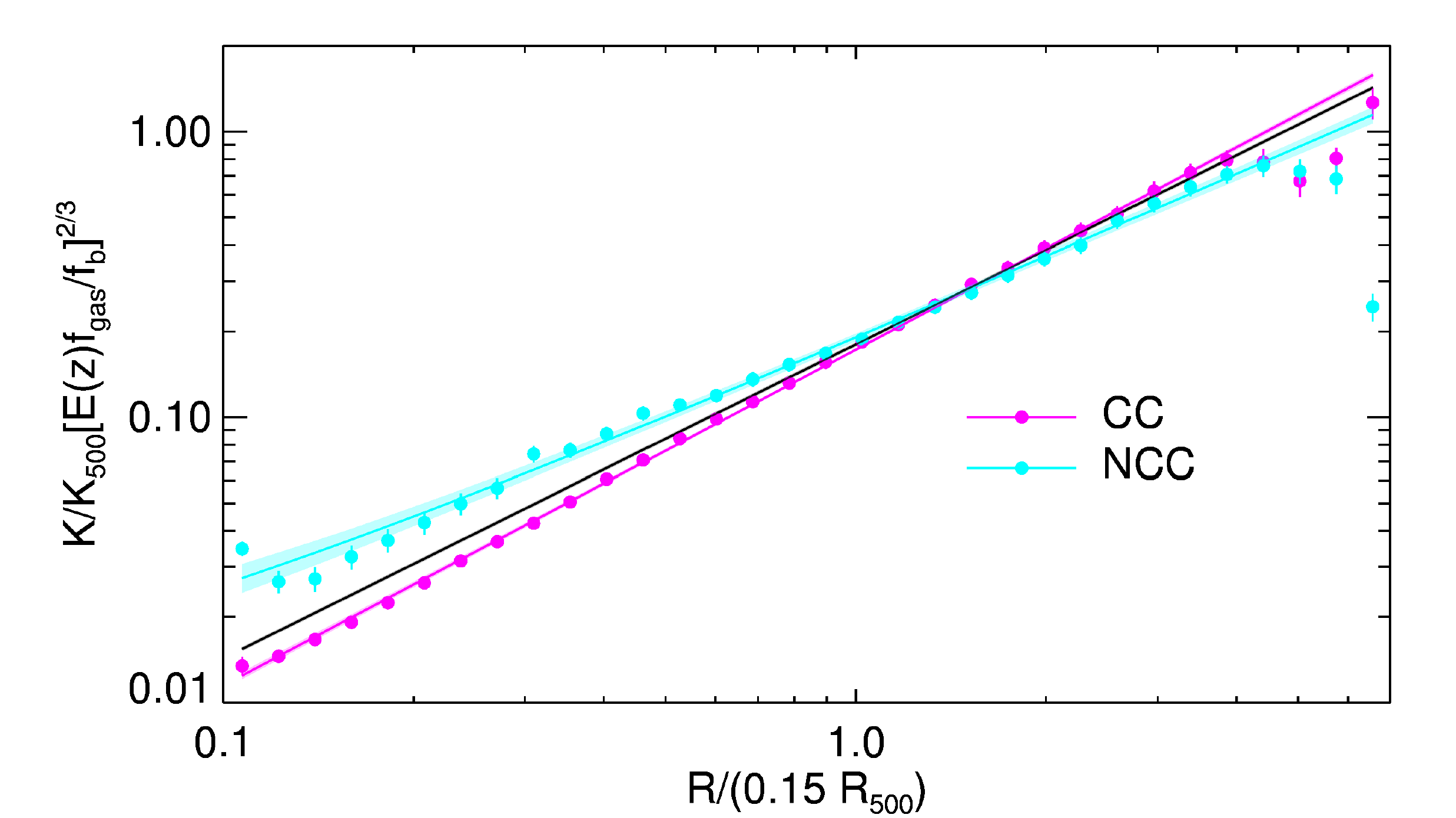}
\caption{Stacked entropy profiles corrected using the gas fraction for the subsets of CC and NCC. CC entropy profile still lie very close to the prediction, while NCC have got closer but still flatter than the prediction.}
\label{fig:KfCCNCC}
\end{figure}

\begin{figure}
\includegraphics[width=0.5\textwidth]{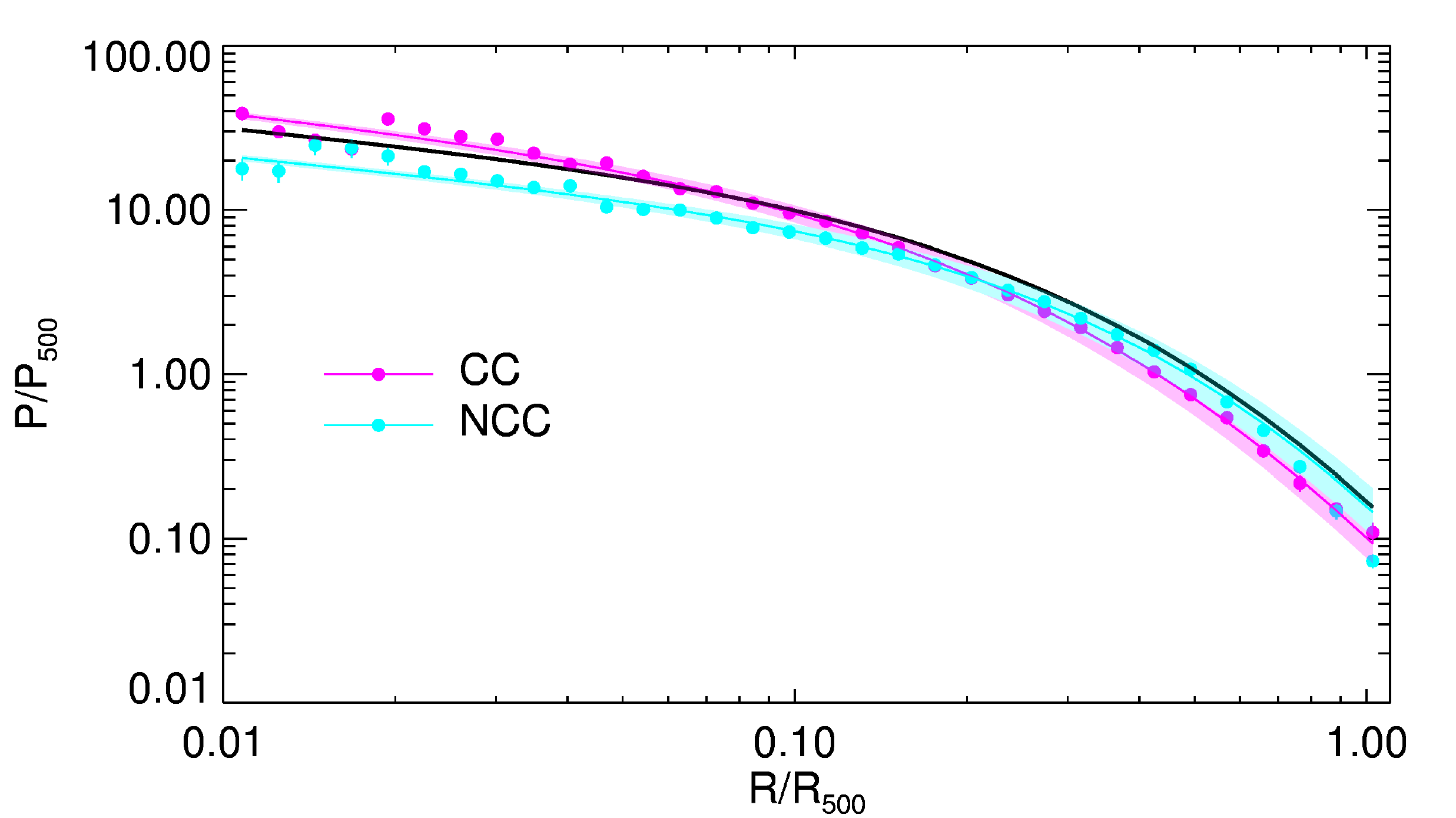}
\caption{Pressure profiles for the subsets of CC and NCC. These profiles are very similar to the one in Fig.~\ref{fig:pressure}, however the NCC profiles within $0.1 R_{500}$ are about a factor of 2 higher than the high redshift one.}
\label{fig:PCCNCC}
\end{figure}

\newpage

\section{Plots}
\begin{figure}[h]
\includegraphics[width=0.5\textwidth]{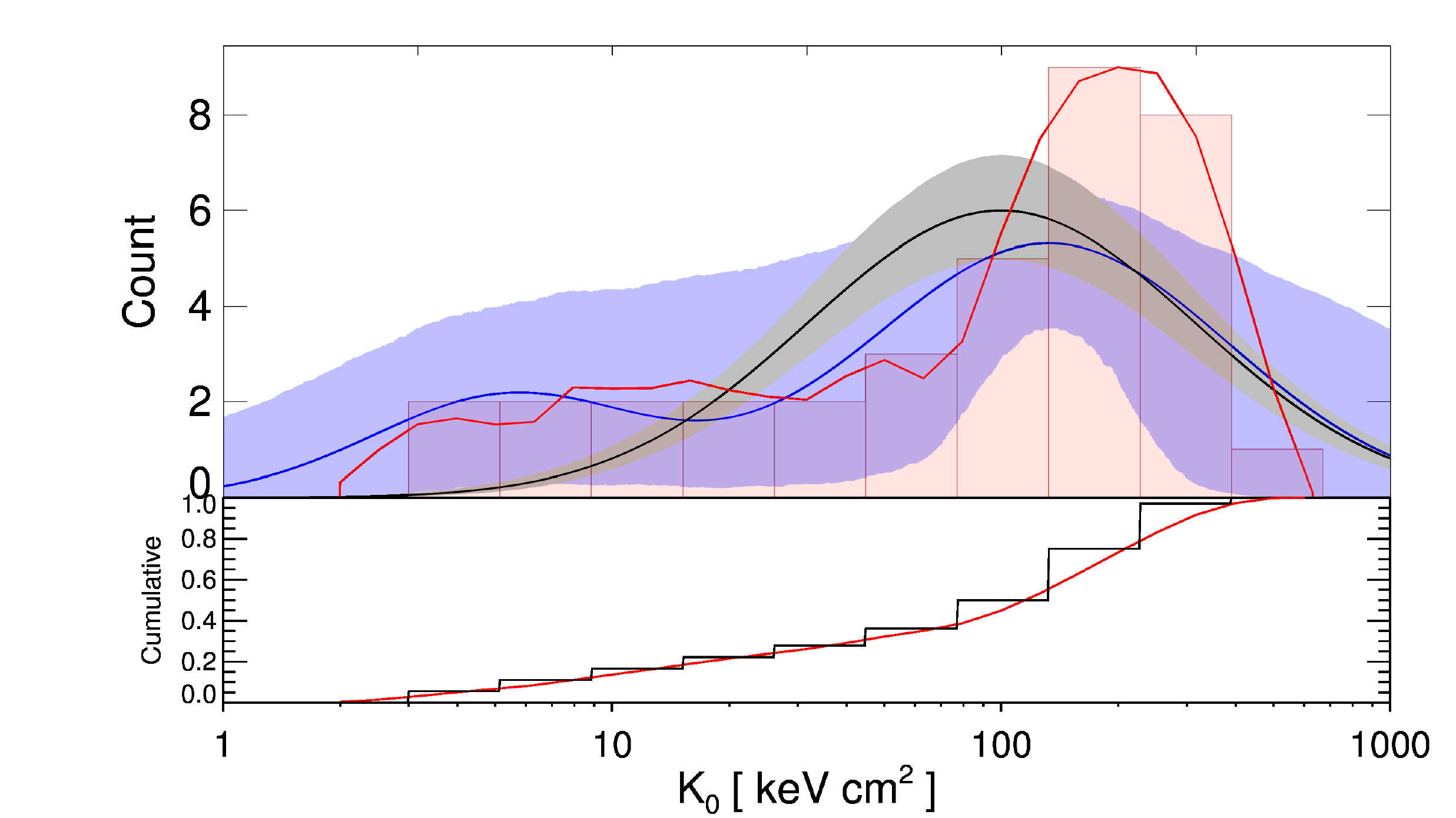}
\includegraphics[width=0.5\textwidth]{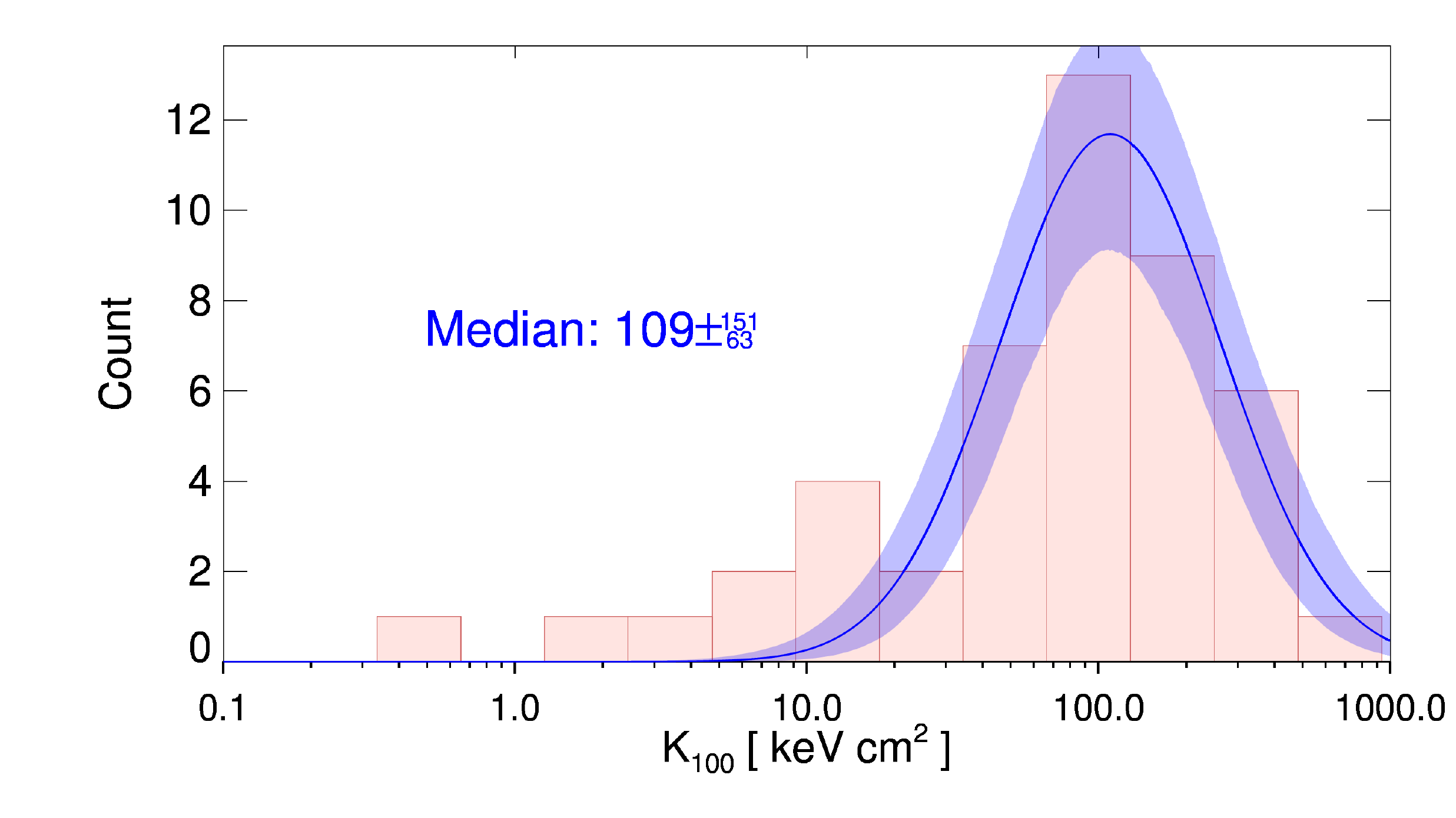}
\includegraphics[width=0.5\textwidth]{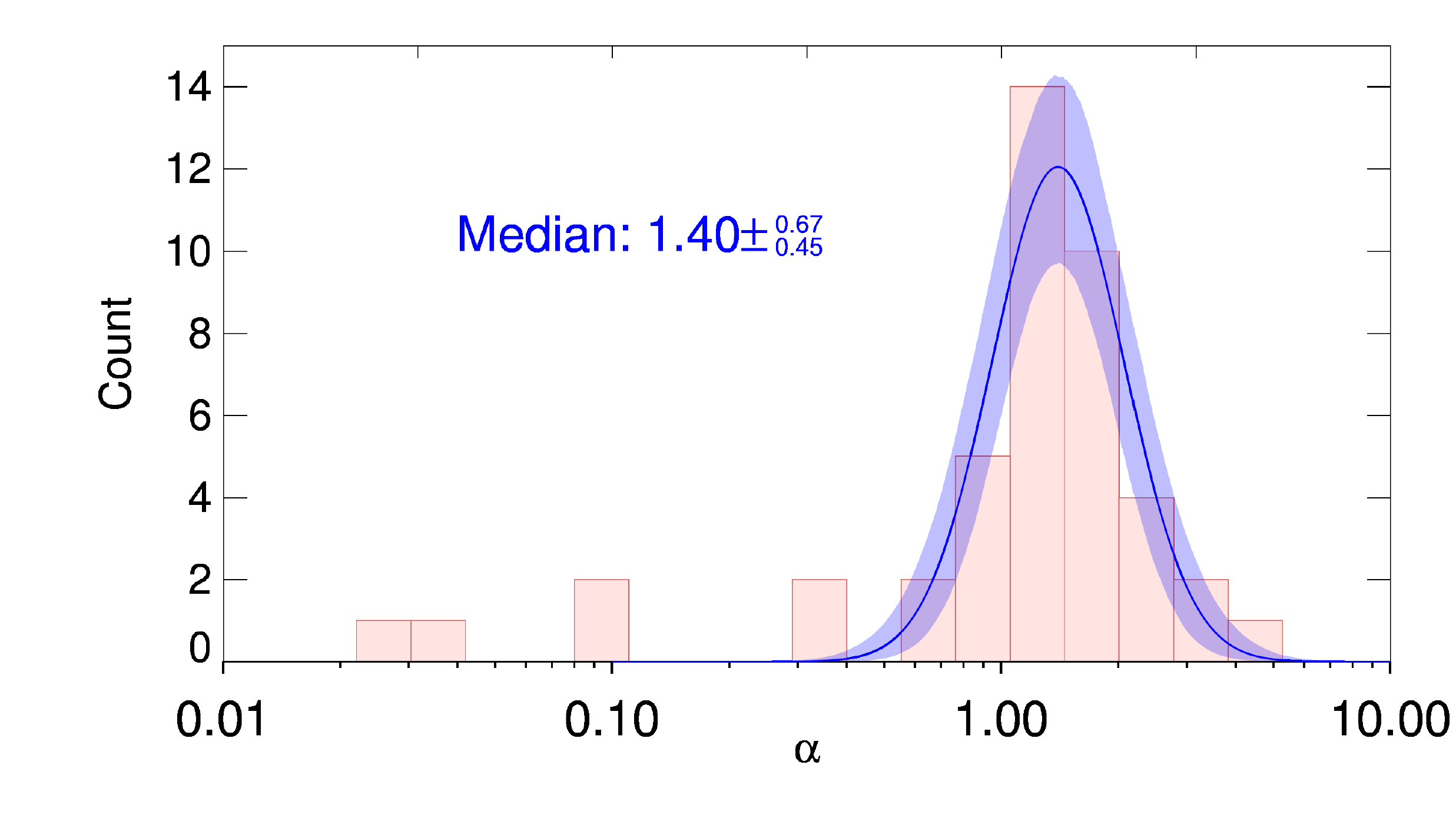}
\caption{
(Top) Distribution of the parameter $K_0$. The red line represents the kernel density plot with a smoothing width of $0.1$. The black and blue line represens the best fit obtained using one and two lognormal distributions respectively. The shaded grey and blue regions enclose the 68.3 \% probability region around the best fit due to parameter uncertainties. (Middle) and (Bottom) Distribution of the parameters $K_{100}$ and $\alpha$ respectively. The blue line with shaded area represents the best fit with a single lognormal distribution and the  1$\sigma$ probability regions around it. The median value of the fit is shown directly on the graph.
}
\label{fig:k0_k100_alpha}
\end{figure}

\begin{figure}[t]
\includegraphics[width=0.5\textwidth]{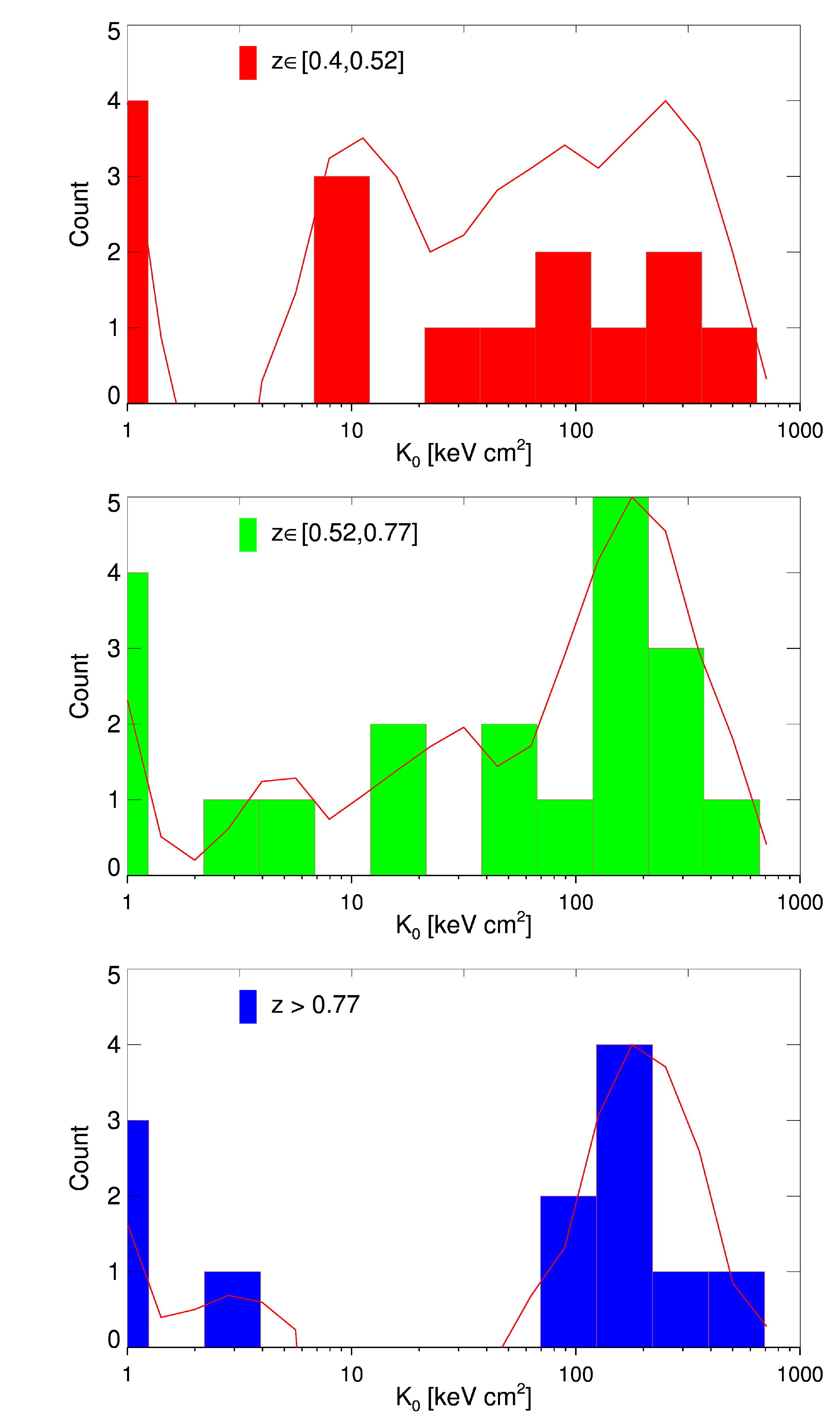}
\caption{Evolution of the central entropy distribution with redshift. We can clearly see an evolution with evidence of bimodality at high redshift. The red lines represents the kernel density estimation of the distributions. The clusters which are best fitted by $K_0 = 0$ are added as if they have the value of 1 kev cm$^2$.}
\label{fig:bimodality_evolution}
\end{figure}

\begin{figure}[t!]
\includegraphics[width=0.5\textwidth]{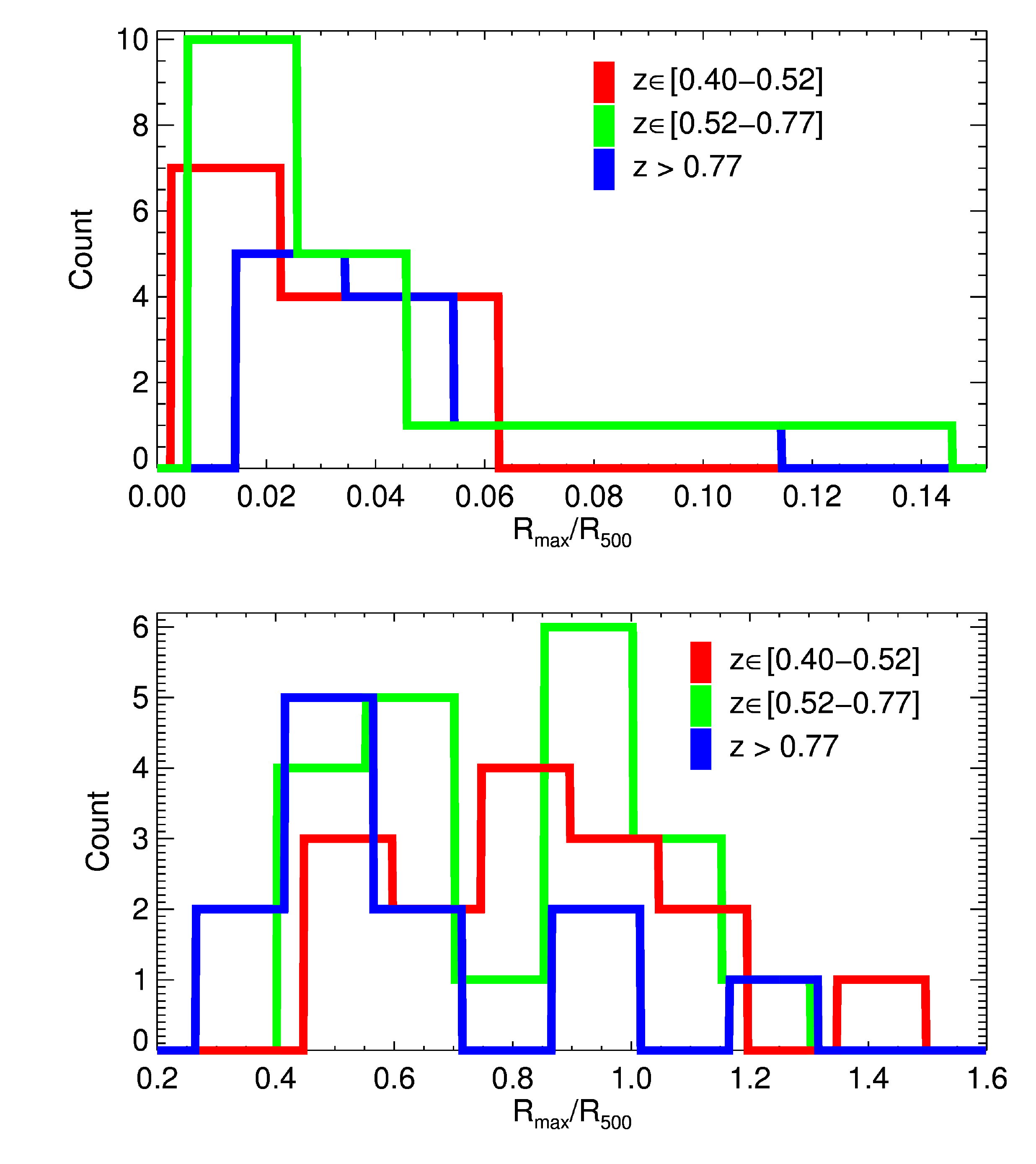}
\caption{
Distribution of the rescaled innermost (Top) and outermost (Bottom) radial spatial bin color coded with redshift.
}
 
\label{fig:maxT}
\end{figure} 

\begin{figure}[h]
\includegraphics[width=0.5\textwidth]{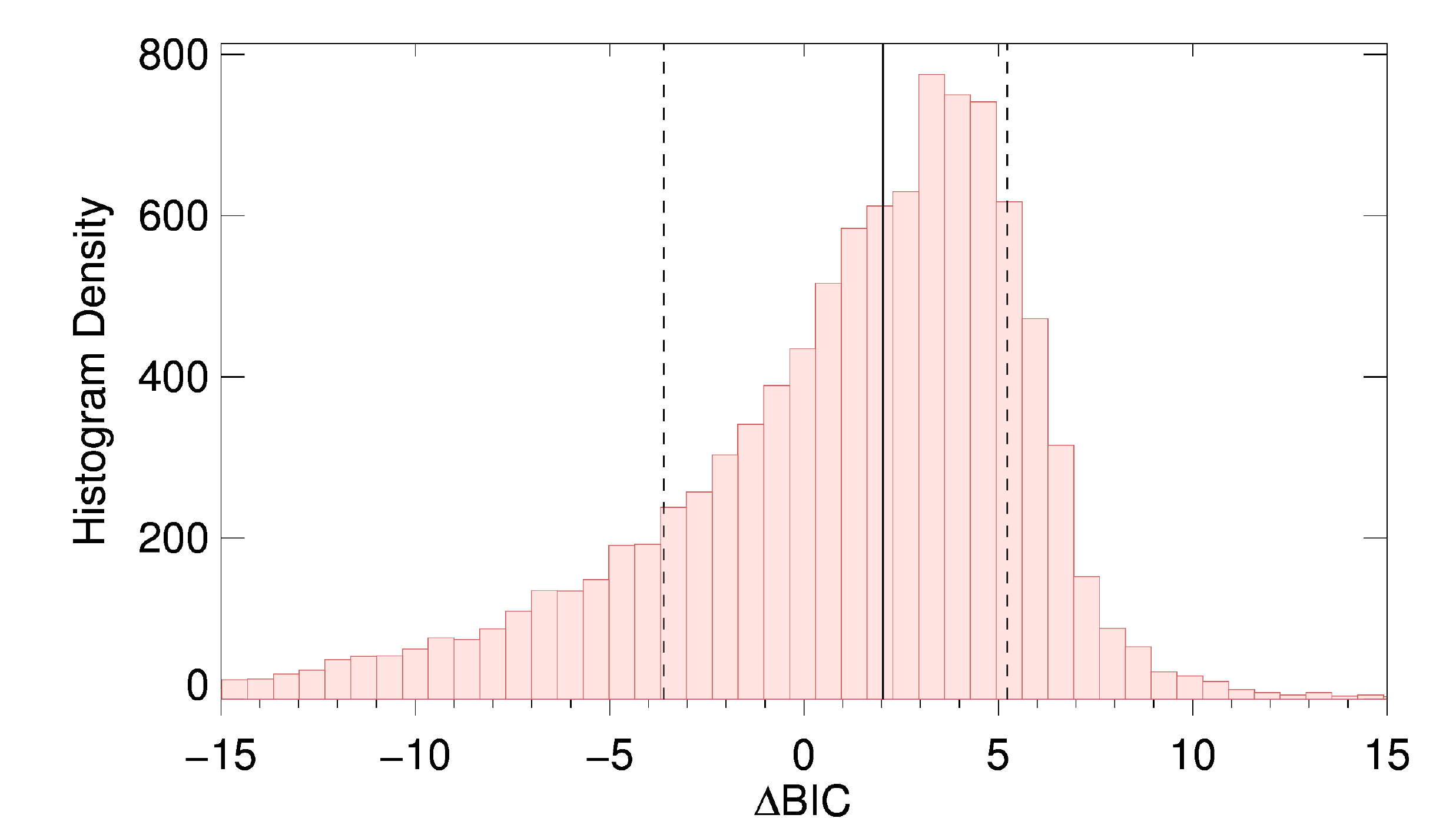}
\caption{Distribution of the bootstraps results. The black solid line represents the position of the median, while the dashed lines represent the region comprehending 68\% of the distribution. The 1$\sigma$ region is compatible with $\Delta$BIC=0. }
\label{fig:BIC}
\end{figure}

\begin{figure}[t]

\includegraphics[width=0.5\textwidth]{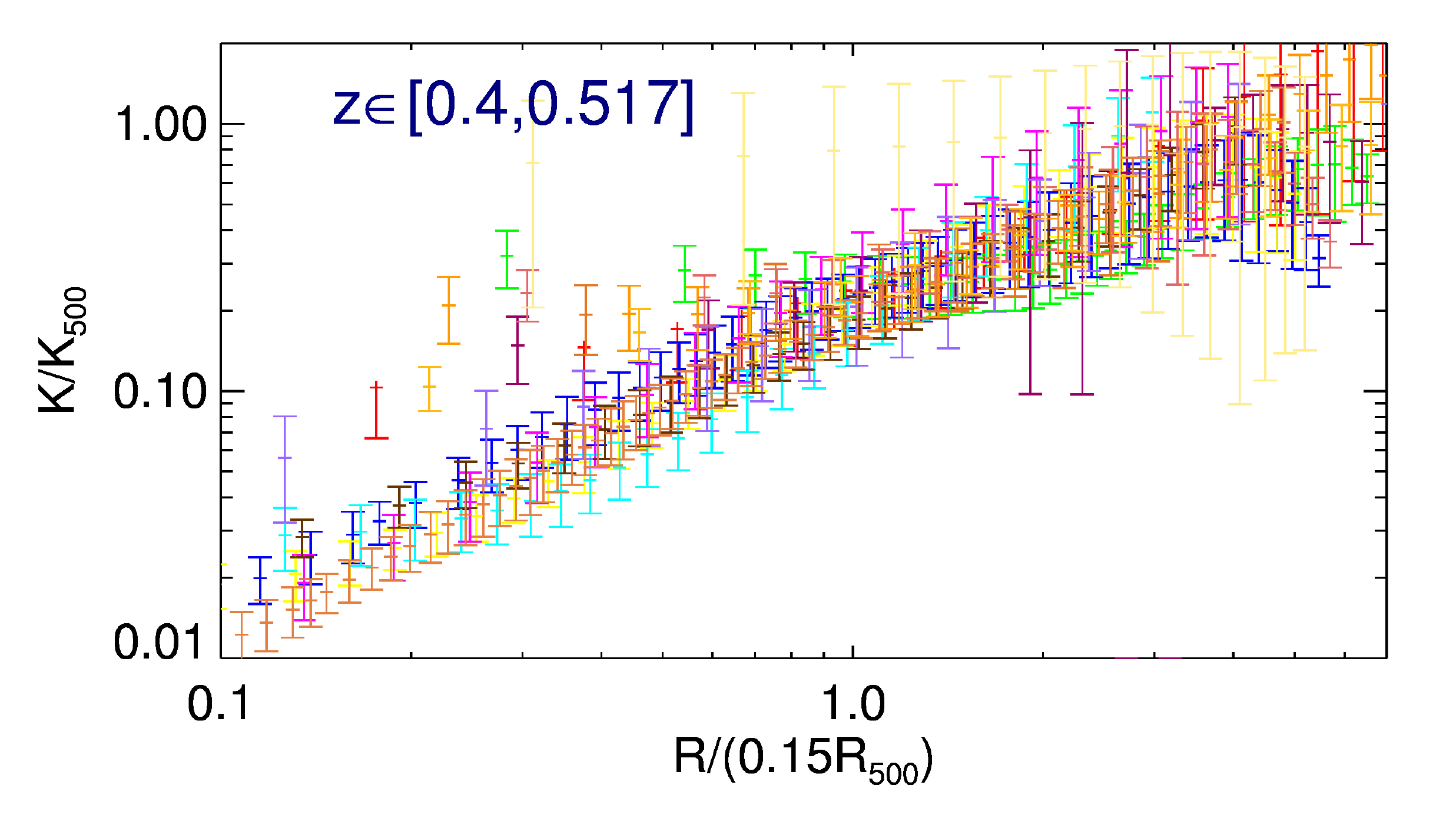}
\includegraphics[width=0.5\textwidth]{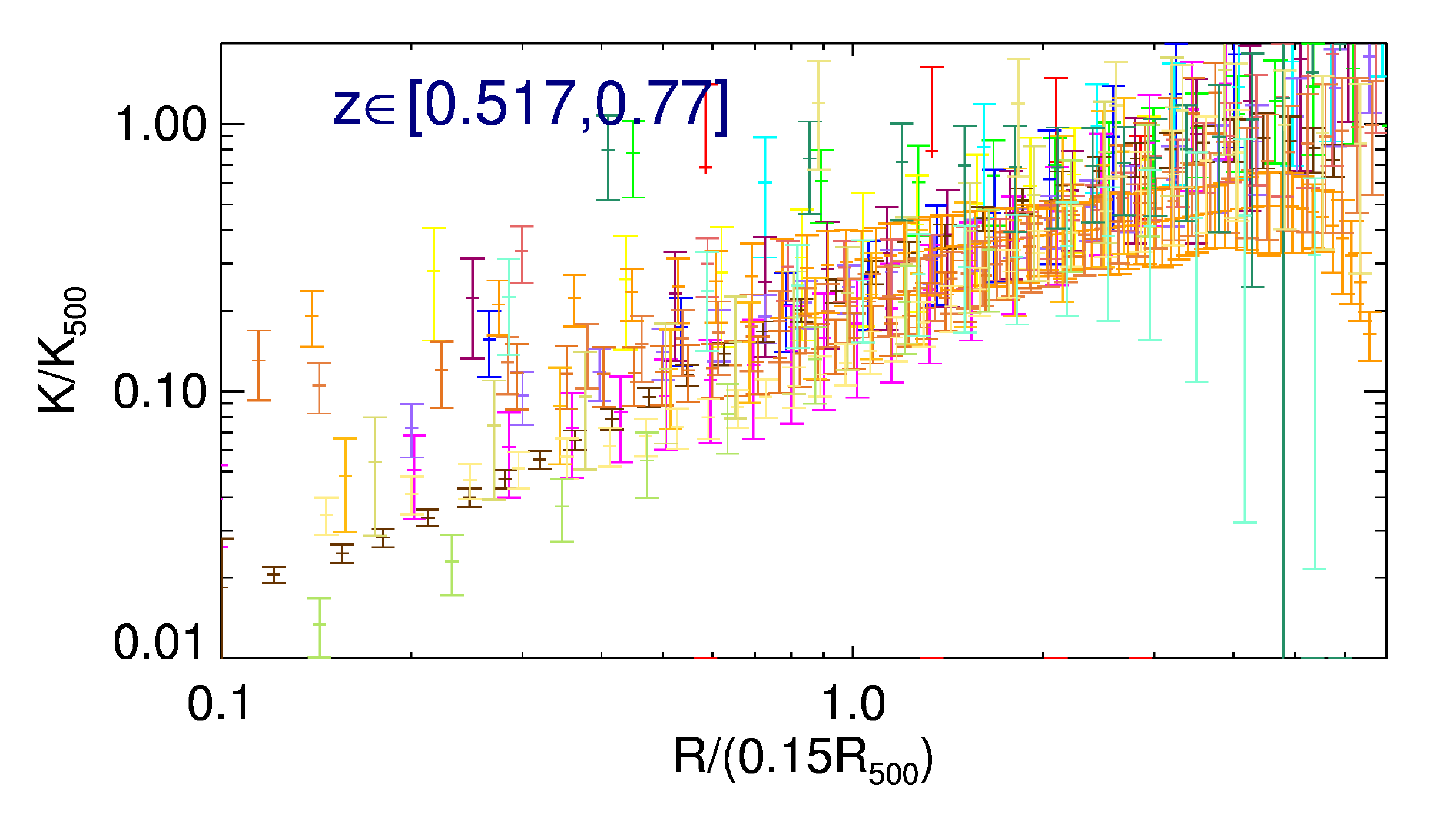}
\includegraphics[width=0.5\textwidth]{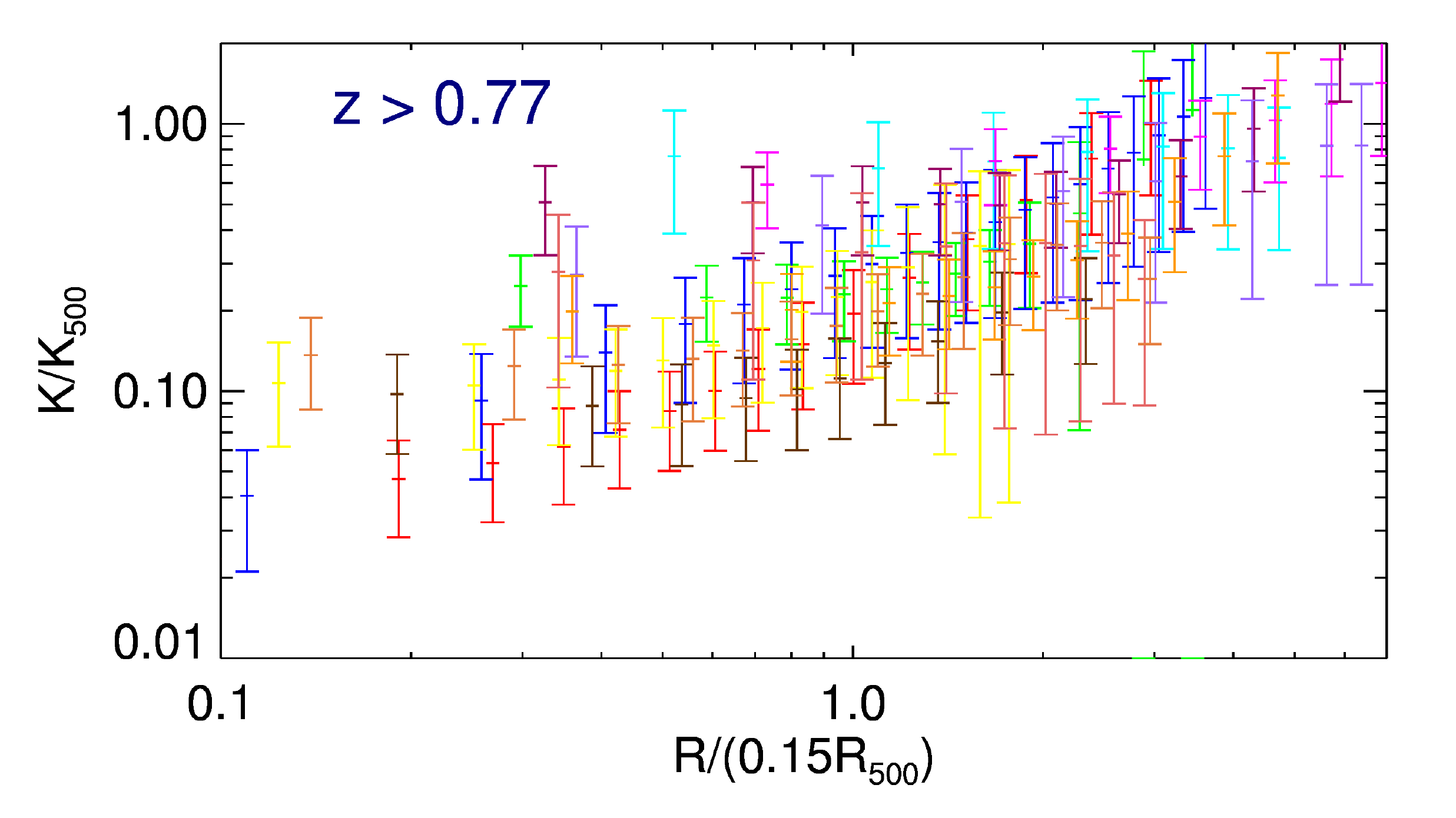}

\caption{Rescaled entropy profiles of clusters in three redshift bins. Each color represents data from a single cluster.}
\label{fig:alldata}
\end{figure}

\end{appendix}

\end{document}